\documentclass[useAMS,usegraphicx,usenatbib]{mn2e}
\usepackage{amsbsy,amsmath,amssymb,amstext,longtable,times,xspace}
\usepackage{natbib}
\usepackage[T1]{fontenc}
\usepackage{aecompl}

\title[The conjectured \nuo retrograde planet]{The conjectured S-type 
retrograde planet in \mbox{$\nu$ Octantis}: more evidence including four years 
of iodine-cell radial velocities}

\author[Ramm \oo]{$\rm D.~J.~Ramm^1$\thanks{E-mail: djr1817@gmail.com}, 
$\rm B.~E.~Nelson^2$, 
$\rm M.~Endl^3$, $\rm J.~B.~Hearnshaw^1$, $\rm R.~A.~Wittenmyer^4$ 
\and $\rm F.~Gunn^1$, $\rm C.~Bergmann^1$, 
$\rm P.~Kilmartin^1$, $\rm E.~Brogt^5$\\
$\rm ^1$ Department of Physics and Astronomy, University of Canterbury, 
Private Bag 4800, Christchurch 8140, New Zealand\\
$\rm ^2$ Center for Interdisciplinary Exploration and Research in Astrophysics 
(CIERA) and Department of Physics and Astrophysics, \\
Northwestern Institute for Complex Systems (NICO), Northwestern University, 
Evanston, IL 60208, USA \\
$\rm ^3$ McDonald Observatory, The University of Texas at Austin, Austin, TX 
78712, USA \\
$\rm ^4$ Department of Astrophysics and Optics, School of Physics, University 
of New South Wales, Sydney 2052, Australia \\
$\rm ^5$ Academic Services Group, University of Canterbury, Private Bag 4800, 
Christchurch 8140, New Zealand \\
}

\date{Released}
\pagerange{\pageref{firstpage}--\pageref{lastpage}} \pubyear{2016}

\newcommand*{\nc}{\newcommand*}

\nc{\esm}{\ensuremath}
\nc{\xs}{\xspace}
\nc{\oo}{et~al.\xs}

\nc{\drb}{Ramm \oo (2009)\xs}
\nc{\nuo}{\esm{\nu~{\rm Oct}}\xs}
\nc{\se}{\section}
\nc{\su}{\subsection}
\nc{\suu}{\subsubsection}
\nc{\noi}{\\ \noindent}

\nc{\sss}{\rm \scriptscriptstyle}
\nc{\lab}{\label}

\nc{\ml}{\multicolumn}

\nc{\wide}{\hspace{1em}}
\nc{\dgap}{\hspace{2em}}
\nc{\hgap}{\hspace{-2em}}
\nc{\ngap}{\hspace{-6em}}

\nc{\mc}{\mathcal}
\nc{\cp}{\esm{\mc{M}_1}\xs}
\nc{\cs}{\esm{\mc{M}_2}\xs}

\nc{\mpl}{\esm{\mc{M}_{\sss pl}}\xs}
\nc{\mjup}{\esm{\mc{M}_{\rm Jup}}\xs}
\nc{\kone}{\esm{\rm 1k\!\times\!1k\ \xs}}
\nc{\kfour}{\esm{\rm 4k\!\times\!4k\ \xs}}
\nc{\vep}{\esm{\varepsilon}\xs}
\nc{\prot}{\esm{P_{\rm rot}}\xs}
\nc{\omp}{\esm{\omega_{_1}}\xs}
\nc{\fmorb}{\esm {f_{\mc{M}_1}}\xs}
\nc{\om}{\esm{\omega}\xs}
\nc{\dg}{\esm{\degr}\xs}

\nc{\mass}{\esm{\mc M}\xs}
\nc{\teff}{\esm{T_{\rm eff}}\xs}
\nc{\mbv}{\esm{M_{\rm V}}\xs}
\nc{\msun}{\esm{~\mc{M}_{\odot}}\xs}
\nc{\rsun}{\esm{~\mc{R}_{\odot}}\xs}
\nc{\mldr}{\esm{_{\sss MLDR}}\xs}

\nc{\chisq}{\esm{\chi^2}\xs}
\nc{\redchi}{\esm{\chi ^2_{\sss \nu}}\xs}
\nc{\sqrchi}{\esm{(\redchi)^{1/2}}\xs}
\nc{\ld}{\esm{\lambda}\xs}
\nc{\iod}{\esm{\rm I_2}\xs}
\nc{\viod}{\esm{V_{\sss I_2}\xs}}
\nc{\vccf}{\esm{V_{\sss CCF}\xs}}

\nc{\tz}{\esm {T_{\sss 0}}\xs}
\nc{\dbin}{\esm{_{\sss bin}}}
\nc{\dpl}{\esm{_{\sss pl}}}
\nc{\apl}{\esm {a\dpl}\xs}
\nc{\epl}{\esm {e\dpl}\xs}
\nc{\abin}{\esm {a\dbin}\xs}
\nc{\ibin}{\esm{i\dbin}\xs}
\nc{\ipl}{\esm{i\dpl}\xs}

\nc{\kms}{\esm{~\rm km\,s^{-1}}\xs}
\nc{\ms}{\esm{~\rm m\,s^{-1}}\xs}
\nc{\skms}{\esm{\,\rm km\,s^{\sss -1}}\xs}
\nc{\sms}{\esm{\,\rm m\,s^{\sss -1}}\xs}
\nc{\hc}{HERCULES\xs}
\nc{\caii}{\mbox{\rm Ca \footnotesize II~}}

\nc{\hip}{{\it Hipparcos}\xs}

\nc{\ie}{i.e.~}
\nc{\eg}{e.g.~}
\nc{\fn}{\footnote}

\nc{\bc}{\begin{center}}
\nc{\ec}{\end{center}}
\nc{\bte}{\begin{table}}
\nc{\ete}{\end{table}}
\nc{\btr}{\begin{tabular}}
\nc{\etr}{\end{tabular}}
\nc{\bfi}{\begin{figure}}
\nc{\efi}{\end{figure}}

\nc{\beq}{\begin{equation}}
\nc{\eeq}{\end{equation}}
\nc{\ca}{\caption}
\nc{\eql}{Eq.~\eqref}
\nc{\fl}{Fig.~\ref}
\nc{\tl}{Table~\ref}
\nc{\fnl}{Footnote~\ref}
\nc{\scl}{\S~\ref}

\nc{\si}{\esm{\sigma}\xs}
\nc{\er}{\esm{\pm}\xs}
\nc{\spm}{\esm{\,\pm\,}\xs}

\nc{\pl}{\esm{_{\rm pl}}}
\nc{\di}{\esm{_{\rm i}}}
\nc{\dgs}{\esm{_{\rm gs}}}
\nc{\dn}{\esm{_{\rm n}}}

\nc{\pp}{^{\prime\prime}}
\def\upp{\hbox{$\stackrel{\lower0.4ex\hbox{$\pp$}}{\lower.05ex\hbox{.}}$}}
\nc{\updeg}{\esm{\hbox{$\stackrel{\lower0.4ex\hbox{\degr}}{\lower0.05ex\hbox{.}}$}}}

\nc{\RUNDMC}{{\textrm RUN\,DMC}\xspace}
\nc{\MERCURY}{{\textrm MERCURY}\xspace}
\nc{\SWARMNG}{{\textrm Swarm-NG}\xspace}

\begin{document}
\lab{firstpage}

\maketitle

\begin{abstract}
We report 1212 radial-velocity (RV) measurements obtained in the years 
2009-2013 using an iodine cell for the spectroscopic binary $\nu$~Octantis 
(K1III/IV). This system ($\abin\sim$~2.6 au, $P\sim1050$~days) is 
conjectured to have a Jovian planet with a semi-major axis half that of the 
binary host. The extreme geometry only permits long-term stability if the 
planet is in a retrograde 
orbit. Whilst the reality of the planet ($P\sim415$~days) remains uncertain, 
other scenarios (stellar variability or apsidal motion caused by a yet 
unobserved third star) continue to appear substantially less credible based on 
CCF bisectors, line-depth ratios and many other independent details. If this 
evidence is validated but the planet is disproved, the claims of other planets 
using RVs will be seriously challenged.

We also describe a significant revision to the previously published RVs and 
the full set of 1437 RVs now encompasses nearly 13 years. The sensitive 
orbital dynamics allow us to constrain the 3D architecture with a broad prior 
probability distribution on 
the mutual inclination, which with posterior samples obtained from an $N$-body 
Markov chain Monte Carlo is found to be $152\updeg5\pm^{0.7}_{0.6}$. None of 
these samples are dynamically stable beyond $10^6$ years. However, a grid 
search around the best-fitting solution 
finds a region that has many models stable for $10^7$ years, and 
includes one model within 1-sigma that is stable for at least 
$10^8$ years. The planet's exceptional nature demands robust independent 
verification and makes the theoretical understanding of its formation a 
worthy challenge.

\end{abstract}


\se{Introduction}
\lab{intro}

It has been speculated from several years of radial velocity (RV) observations 
that the single-lined spectroscopic binary (SB1) $\nu$ Octantis hosts a 
Jovian planet (Ramm 2004; Ramm \oo 2009). Given the present scale of exoplanet 
discoveries, this may not seem too unusual. However, the binary has by far the 
tightest geometry of any proposed to harbour a circumstellar (\ie S-type) 
planet, having a mean separation of only $\abin\sim2.6$~au. The 
minimum-separation barrier is presently tentatively at about 20~au, as will be 
described in more detail shortly. Furthermore, the RV signal, if it is 
caused by a planet, places its circumprimary orbit about midway between the 
stars \ie $\apl\sim0.5\abin$. For a prograde orbit, models predict long-term 
stability should be limited to about half this distance \ie 
$\apl\lesssim0.25\abin$ (Holman \& Wiegert 1999; Ramm \oo 2009; Eberle \& 
Cuntz 2010; Andrade-Ines \oo 2016). However if the planet is in a retrograde 
orbit, as first suggested by Eberle \& Cuntz (2010), the stability zone is 
wider (Jefferys 1974; Morais \& Giuppone 2012), and places the conjectured 
orbit very near and perhaps within the boundary of stability. As extraordinary 
as the retrograde scenario may appear, since its very formation would also be 
challenged 
by strong dynamical interactions, the alternative standard explanations of 
instrumental or data reduction anomalies, starspots, pulsations or apsidal 
motion of the binary orbit are less supported by the observational evidence 
(Ramm \oo 2009; Ramm 2015). 
Other retrograde planets have been postulated but in each of them 
a choice exists between viable prograde and retrograde options (see \eg 
Gayon-Markt \& Bois 2009; Go{\'z}dziewski \oo 2015) whilst no choice exists for \nuo.

Nu Octantis (HD\,205478 HIP\,107089 HR\,8254) has a slightly evolved early 
K-type primary with a late-dwarf or white-dwarf secondary (\tl{stellar}). 
The possibility of a planet was first mentioned in Ramm (2004) with 156 RVs 
and barely one binary orbit observed. The planet elements were 
approximated to $P\sim400$~days, $e=0$, $K\sim60\ms$ and 
$\mpl\sin\ipl\sim3\mjup$. The 
tight orbital geometry and commonly found rotation period of evolved K-type 
stars led Ramm to surmise, though mostly on qualitative evidence, that 
rotational modulation of recent surface features was a more likely cause. 
With three more years of data, \drb identified the planetary hypothesis as 
having more support, this explanation though very precarious given the rapid 
instability of a prograde orbit. They claimed the then considered alternatives 
had less 
credibility because: (1)~\hip provided evidence of photometric 
stability ($H_p=3.8981\pm0.0004$; ESA 1997), (2)~there was no correlation of 
bisector values with the perturbation RVs, (3)~the anticipated upper bounds of 
the rotation period and the need for unusually long-lived surface features, 
and (4)~there was no evidence of significant activity, chromospheric or 
otherwise (Warner 1969; Slee \oo 1989; Beasley, Stewart \& Carter 1992; 
H\"{u}nsch \oo 1996). Morais \& Correia (2012) suggested the 
planet may be an illusion created by the system being instead a hierarchical 
triple, with \nuo~B being a binary and causing apsidal motion of the observed 
primary's orbit. This alternative, however, has no support from the already 
published orbits spanning 100 years (Ramm 2004; Ramm \oo 2009).

The first orbital solutions for the conjectured system (Ramm \oo 2009) are 
summarised in \tl{orbital}. Combined with astrometry from \hip (ESA 1997), 
the longitude of the line of nodes $\Omega\dbin$ and the inclination $i\dbin$ 
were also reported, the latter being one of the most precise estimates derived 
using this method. This preliminary solution, based on the understandable 
uncertainty of the underlying cause, was determined assuming only a simple 
double-Keplerian model. 

\bte
\bc
\btr{lcc}
\hline
  Parameter       &  $\nu$ Octantis A               & Reference \\ 
\hline
Spectral type     &   K0III $\ddag$               & (3) \\
$V$ (mag)         &  $3.743\pm0.015$              & (1) \\ 
$\mbv$ (mag)      &  $+2.02\pm0.02$               & (2) \\ 
$(B-V)$           &  $0.992\pm0.004$              & (1) \\ 
$H_{\rm p}$ (\hip mag)  &  $3.8981\pm0.0004$      & (3) \\ 
Mass (\msun)      &  1.61                         &  (6) \\ 
Radius (\rsun)    &  $5.81\pm0.12$                &  (6) \\ 
\teff (K)         &  $4\,860\pm40$                &  (6) \\ 
$\log g$ (cgs)    &  $3.12\pm0.10$                &  (6) \\ 
$\rm [Fe/H] (dex)$&  $+0.18\pm0.04$               & (6) \\ 
$v\sin i$ (\kms)  &  2.0                          & (4),(6) \\ 
Age (Gyr)         &  $\sim2.5$--3                 & (5) \\         
\hline							                     
\etr
\ca{Stellar parameters for \nuo. 1:\,Mermilliod (1991), 
2:\,present work, 3:\,ESA (1997), 4:\,Costa \oo (2002); 5:\,Ramm \oo (2009), 
6:\,Fuhrmann \& Chini (2012). Fuhrmann \& Chini revised the primary mass and 
radius estimates given in \drb, based in particular on their metallicity. 
Their 2\si errors are halved here to be consistent with the 1\si errors used 
elsewhere here. $\ddag$ We suspect the luminosity class may be evolved at most 
IIIb - IV (see the Discussion).}
\lab{stellar}
\ec
\ete

\bte
\bc
\btr{lcc}
\hline
                  & \ml{2}{c}{\nuo A's absolute orbits} \\
\hline
companion         & \nuo B               & conjectured planet \\   
$K_1$ (\kms)      & $7.032\pm0.003$      & $0.052\pm0.002$    \\ 
$P$ (days)        &  $1050.1\pm0.1$      &  $417.4\pm3.8$     \\ 
$e$               &  $0.2359\pm0.0003$   &  $0.12\pm0.04$     \\ 
\omp (\dg)        & $75.05\pm0.05$       &  $260\pm21$        \\ 
$i$ (\dg)         &  $70.8\pm 0.9$       &  $-$               \\ 
$\Omega$ (\dg)    &  $87\pm 1.2$         &  $-$               \\ 
${\mc M}\sin i$   &  $0.55$ (\msun)      &  $2.4$ (\mjup)     \\ 
$a$  (au)         &  $2.6\pm0.1$         &  $1.3\pm0.1$       \\  
RMS (\ms)         &  \ml{2}{c}{19}                      \\
\sqrchi           &  \ml{2}{c}{4.2}                \\
\hline
\etr
\ca{Orbital parameters for \nuo~A and the conjectured planet from 222 RVs 
(Ramm \oo 2009) and \hip astrometry. The 
Keplerian elements are the RV semi-amplitude $K$, period $P$, eccentricity 
$e$, the argument of periastron \om. The scaled masses are based on the 
revised primary mass. Each relative orbit's semimajor axis $a$ is derived from 
Kepler's Third Law, and assumes coplanar orbits.}
\lab{orbital}
\ec
\ete

The retrograde-orbit scenario for \nuo has been explored in increasing 
complexity by Quarles, Cuntz \& Musielak (2012) and Go{\'z}dziewski \oo 
(2013). Both studies conclude that stable retrograde orbits exist that obey 
the rather noisy observational constraints. As well as being the first to 
re-analyse the RVs with a self-consistent Newtonian model, Go{\'z}dziewski \oo 
(2013) also emphasised the complex chaotic space of narrow intersecting 
stability 
zones and relatively small islands of long-term stability. Furthermore, their 
stable models had to be significantly shifted from the $N$-body and Keplerian 
best-fit models, implying the conjectured planet was inconsistent with the 
then available data.

Planets may exist in binary systems in either P-type (in the present context,
circumbinary) or S-type (circumstellar) configurations. Of the confirmed 
binaries harbouring S-type planets, all have stellar separations exceeding 
that of \nuo by an order of magnitude or so. Zhou \oo (2012) and  Wang \oo 
(2014) identify the apparent 
barrier of about 20~au to such orbits based on the accepted planet-hosting 
binaries. These include GJ~86~A (Queloz \oo 2000), HD\,41004~A 
(Zucker \oo 2004), and HD\,196885~A (Correia \oo 2008; Th{\'e}bault 2011). 
One of the first exoplanet claims was for $\gamma$~Cephei~A (Campbell, Walker 
\& Yang 1988) which was then understandably challenged (Walker \oo 1992). 
Eleven years elapsed before Hatzes \oo (2003) provided evidence that 
vindicated that planet's reality. An S-type planet was 
claimed for $\alpha$~Cen~B (Dumusque \oo 2012; $\abin\sim17.5$~au) though the 
planet orbit is tiny ($\apl=0.04$~au). Its existence has been 
challenged (Hatzes 2013) and more recently by Rajpaul, Aigrain \& 
Roberts (2016). The planet claimed for the hierachical triple HD~188753 
(Konacki 2005; $\abin\sim13$~au) also appears to be non-existent 
(Eggenberger \oo 2007). 
All accepted S-type systems have separation ratios much smaller than 
that implied for \nuo, with $\gamma$~Cep, HD\,196885 and HD\,41004 all having 
$\apl/\abin\sim0.1$. It is thus hardly surprising that the \nuo planet remains 
unconfirmed and controversial,\fn{Approximately 10\% ($\sim200$) of present 
planet claims are in this category. See The Extrasolar Planets Encyclopeadia 
at http://exoplanet.eu} even though it may so far `pass' many standard tests 
for such a claim, as the above-mentioned discredited planets also originally 
did.

Thus there is no direct observational precedent for a planet to exist in such 
circumstances though, for instance Trilling \oo (2007) gives some indirect 
evidence based on infra-red excesses of main-sequence binaries. The \nuo 
system also presents formidable theoretical 
challenges and many papers explore the formation and stability of S-type 
systems, such as Chauvin \oo (2011), Th{\'e}bault (2011), M{\"u}ller \& Kley 
(2012), Jang-Condell (2015), and Rafikov \& Silsbee (2015). The planet forming 
coevally with the present binary orbit is inconsistent with 
our present understanding of such processes, even if the original orbit was 
prograde. Rafikov (2013) studied planetesimal formation in 
small-separation binaries (again at the `barrier' $\abin\sim20$~au) and found 
that sufficiently massive axisymmetric disks may eliminate the fragmentation 
barrier for the formation of giant planets in these `tight' systems. Whether 
or not these findings or related ones can ultimately extend to the extreme 
demands of \nuo remains to be determined. Hence, it is not surprising that 
such tight binaries are rarely included in S-type planet search programmes, 
and this bias may underlie (though surely unexpectedly) their lack of 
discovery.

Since coeval formation is presently so unbelievable, more credible 
alternatives must be considered, no matter how exotic they may presently seem. 
A strong gravitational event (\eg a passing star) is not so extreme. This 
multi-faceted possibility includes variations such as capture of one other 
star or a planet-hosting star. Dynamical  captures or 
exchanges have been discussed by Pfahl \& Muterspaugh (2006), and in the 
context of open cluster interactions by Portegies Zwart \& McMillan (2005). 
Another variation is that the planet was originally hosted by the secondary 
and somehow exchanged (see \eg Kratter \& Perets 2012; not surprisingly, their 
work focussed on more manageable binary models having \eg separations of 
75--105 au). Tutukov \& Fedorova (2012) describe other atypical planet 
formation scenarios, including one conjectured to yield retrograde orbits 
by passage of the host through a dense rotating molecular cloud.

This paper continues the search for the cause of the RV behaviour of \nuo. In 
Section 2 we describe the nearly 13 years of data and the two instrument 
setups, the second including a larger detector and iodine cell. Section 3 
presents our RV datasets, including the new iodine-cell velocities acquired 
between 2009 and 2013, and a critical revision of the older CCF RVs 
(2001--2007) reported in \drb. We also provide preliminary orbital solutions, 
at this point abbreviations of our simple double-Keplerian model. In Section 4 
we assess \caii H lines, cross-correlation function (CCF) bisectors, and 
line-depth ratios. Section 5 provides our 3-D dynamical orbital solutions, and 
our stability modelling results are given in Section 6. Section 7 discusses 
three additional system details: our derived space velocities, the benefits of 
future astrometry and imaging opportunities, and the nature of the system's 
initial discovery. The Discussion evaluates the various possible causes of the 
mysterious RV signal. The paper concludes that the hypothesis, that the RV 
signal is caused by a retrograde planet, still has the most support from the 
observations.

\se{Observations and reductions}
\su{Spectrograph and CCD detectors}
All spectroscopic observations were obtained at University of Canterbury Mt. 
John Observatory, Lake Tekapo, New Zealand using the 1-m McLellan telescope 
and the \'{e}chelle spectrograph \hc (Hearnshaw \oo 
2002; Ramm 2004). From the start of operations in 2001 until 2006/7 
the detector was a \kone detector. Initial testing of a 
complete-wavelength-coverage \kfour detector began in October 2006. This 
detector was then removed briefly and the original detector used until 
the \kfour CCD was installed permanently in early 2007.

\su{Observations}
As well as \nuo's many favourable spectral and photometric characteristics 
for precise RVs, its airmass at Mt.~John (latitude $43\dg$~S) never exceeds 
1.9 due to its far southern declination ($\sim-77\dg$).

\suu{\kone CCD (2001--2007)}
\lab{rejects}
In addition to the 222 observations (2001-2006) reported in \drb, a further 
21 spectra were obtained over four consecutive nights February--March 
2007 between the detector changes described above and extend the time-span of 
observations with this detector to nearly 2100 days. It has also been 
decided to reject eighteen of the Ramm \oo paper's spectra from our 
final analyses as it has been recognised they were probably acquired in 
circumstances making their reliability and quality questionable.\fn{See Ramm 
(2015; \S~2.1.1) for an explanation of these additions and rejections.} 
Thus 225 
RV spectra were included from this detector, 215 with a resolving power 
$R\sim70\,000$ and ten with $R\sim41\,000$. The typical exposure time for 
stellar spectra was 3--5 minutes which provided an average signal-to-noise 
ratio $S/N=122\pm16$ in the middle of order $n=110$ ($\ld\sim5170$~\AA).

\suu{\kfour CCD (2006--2013)}
\lab{blue}
During the brief initial commissioning of this detector in late 2006 (but 
without the \iod cell), two \nuo spectra were taken on the night of October 15 
(JD\,245\,4024). When this larger CCD was installed permanently in early 
2007, the \iod cell was included, positioned just before the Cassegrain 
focus. The cell was maintained at 50\dg~C and provided a dense imprint of 
molecular absorption lines from 5000 to 6200~\AA. This cell 
was replaced in March 2011 by one promising to provide somewhat stronger 
spectral lines.\fn{However, as \tl{newccdiod} will reveal, the mean internal 
RV error for our \nuo spectra actually increased by 0.7\ms with this change.} 
A pinhole relay system was installed in March 2011 in front of the fibre 
entrance. The pinhole was included for the benefit of a 
parallel programme observing $\alpha$~Cen (Endl \oo 2015), whose 
increasingly narrow angular separation contributed significant 
cross-contamination. The pinhole reduced the light throughput by about half 
and eventually (April 2013) it could be removed when required.

All spectra obtained with this detector used $R\sim70\,000$. Stellar exposure 
times were generally 12--20 minutes with the pinhole, and 7--15 minutes 
without it. A total of 1180 spectra had iodine-cell 
RVs derived from them, having an average $S/N=113\pm26$. A further 32 spectra 
could only have a CCF RV derived, and fell into two groups: 
22 RV exposures with $S/N=157\pm44$ and ten \caii -related exposures averaged 
$349\pm49$. Thus there are 1233 new RV spectra reported from both 
detectors.

Initially only spectral orders $n=86-116$ were recorded. From 
December 2010, the readout was extended to $n=129$, beyond 
which inadequate $S/N$ was found for our standard RV exposures. These bluer 
orders promise more precise CCF RVs, and critically, would be needed for our 
CCF bisector measurements. These orders would also allow us to recover RVs for 
\kfour spectra for which \iod-cell processing proved unachievable, and allow 
us to significantly improve the precision of the \kone RVs presented in \drb 
(\scl{i2ccfcal} and \scl{fixk1}).

\su{Preliminary spectral reductions}
\lab{nlambda}
The non-iodine part of each reduction was carried out using the Hercules 
Reduction Software Package HRSP v.3 for the \kone spectra (Skuljan 2004; also 
see Ramm 2004 for improvements to CCF RVs) and HRSP v.5 for the 
\kfour spectra. Standard 
methods of {\'e}chelle spectral reduction were utilised that included 
background subtraction and cosmic ray filtering. A quartz white lamp was used 
for flat fielding and normalization. Th-Ar spectra 
were obtained immediately before and after the stellar exposures for 
wavelength calibration. This latter requirement was critical for the 
CCF-derived RVs, but substantially less important for the iodine-cell RVs, as 
the \iod lines provide the wavelength calibration. An 
interpolated dispersion solution from the thorium spectra was calculated 
corresponding to the flux-weighted-mean observation time, $t_{\rm fwm}$, the 
latter provided by an exposure meter. The $t_{\rm fwm}$ allowed HRSP to 
calculate the Julian-Day and RV barycentric corrections, which are the dates 
and RVs used throughout this paper.

\se{Radial velocities}
The 257 CCF and 1180 \iod radial velocities give a total of 1437 RVs sampling 
a time-span of nearly 4600 days or about 12.5 years. The method for deriving 
the \kfour RVs is given first as it is conventional and straightforward. These 
\iod RVs were also required for determining, by a novel strategy, the 
257 CCF RVs with higher precision than had been previously achievable with \hc.

\su{\kfour detector RVs}
\lab{newccfrvs}
Three high S/N iodine-free template spectra of \nuo were co-added to form a 
single reference spectrum. Analysed in conjunction with high S/N spectra 
of the relatively line-free B-stars $\beta$~Centauri and $\alpha$~Eridani, all 
obtained close in time, this reference was deconvolved from the instrument 
profile using the Maximum Entropy Method. The RVs were derived with the 
pipeline AUSTRAL (Endl, K{\"u}rster \& Els 2000) using fourteen orders. An 
abbreviated list of the final velocities, \viod, is given in \tl{newccdiod}. 
The mean error is $3.6\pm0.5\ms$. We also include our barycentric 
corrections and $S/N$ estimates for order $n=110$.

\bte
\bc
\caption{An abbreviated list of the 1180 AUSTRAL-processed \iod spectra 
acquired during 2009-2013 using the \kfour CCD including barycentric-corrected 
Julian date, corrected relative velocity \viod, internal error $\si\di$ and 
barycentric correction, and the $S/N$ in order $n=110$. The entire table 
is provided online.}
\setlength{\tabcolsep}{4pt}
\btr{crcrc}
\hline
\ml{1}{c}{JD}           & \ml{1}{c}{\viod}   & \ml{1}{c}{$\sigma\di$} & \ml{1}{c}{bary.corr.}  & \ml{1}{c}{$S/N$}      \\
\ml{1}{c}{$245\ldots$}  & \ml{1}{c}{(\skms)}  & \ml{1}{c}{(\sms)}      & \ml{1}{c}{(\skms)}     & \ml{1}{c}{($n=110$)} \\
\hline
   4854.1147 &\wide $   3.925 $ &\wide    3.6 &\wide $   3.5271 $ &    84  \\
   4854.1190 &\wide $   3.926 $ &\wide    3.6 &\wide $   3.5302 $ &    88  \\
   4854.1226 &\wide $   3.923 $ &\wide    3.5 &\wide $   3.5329 $ &    89  \\
   4854.1270 &\wide $   3.931 $ &\wide    3.7 &\wide $   3.5360 $ &    85  \\
   4855.1861 &\wide $   3.958 $ &\wide    4.3 &\wide $   3.8538 $ &    72  \\
 \hline
\etr
\lab{newccdiod}
\ec
\ete

\suu{Using $I_2$ RVs as a gauge to derive CCF RVs}
\lab{i2ccfcal}
Thirty-two \kfour spectra could not be fully processed by AUSTRAL. Instead, 
using the orders blueward of the \iod forest, their RVs were derived using 
CCFs. Here we describe the unconventional strategy used to derive the 
CCF-order weights to yield RVs with accuracies comparable to the \iod RVs.

Our only assumption was that the \iod RVs are a far more accurate gauge than 
typical \hc CCF RVs, easily justified by the vast literature supporting this 
detail, and avoids the assumption of a specific model as Ramm \oo (2009) did. 
A continuous series of 561 \kfour stellar spectra was selected that 
provided \iod RVs and the 16 non-\iod orders $n=114-129$. These covered a 
time-span approximating one binary period of $\sim$~1050~days and thus the full 
range of \nuo RVs. This series began on the first night the required bluer 
orders were recorded (JD\,245\,5534) and ended on JD\,245\,6566. They are 
identified in \fl{allrvs}. A high S/N spectrum was used as the reference to 
create the CCFs for these 16 orders $n$. We used the wider $n\ld$ limits 
available with this larger detector ($\sim 50$~\AA) and a Gaussian for 
fitting. The difference between the \iod RV and the CCF RV for each order was 
calculated for the 561 spectra $i$:
\beq
\Delta V_{\rm i,n}=V_{\rm \sss \iod,i} - V_{\rm \sss CCF,i,n} \ , \nonumber
\eeq
\noindent
and the corresponding standard deviation of these differences 
($\rm \si_{\sss \Delta V}$) calculated, each difference weighted by the error 
on $V_{\rm \sss \iod}$. Our goal was to find the Gaussian that 
provided the smallest average of the sixteen $\rm \si_{\sss \Delta V}$ values. 
This set of Gaussians increased in radius from the CCF core in steps of $\pm1$ 
bin from a minimum $\rm r\dgs=\pm4$ bins, to one that assured us the optimum 
radius had been identified.\fn{Our reduction software HRSP also allowed the 
use of third-order spline interpolation and parabolas of varying radii for CCF 
fitting. None of these provided as optimum a final result as that achieved 
with any Gaussian. The possibility of determining the best Gaussian radius 
separately for each order was not pursued for two reasons: 1.~it would have 
added an extra layer of complexity to this analysis, and 2.~the results we 
will soon describe provided CCF RV precision that seemed difficult to claim 
could be improved.} The standard deviations for each order 
for the optimum radius determined its relative weight: 
$w\dn=1/<\si_{\sss \Delta V}>^2$. Finally, due to their differing 
reference spectra, the \iod and CCF RVs have different velocity zero-points. 
Their offset, incorporated into the final published velocities, $V\di$, is 
$V_{\rm off16} = -3.987\pm0.010$\kms. The total error assigned to each CCF 
$V\di$ was this offset error ($\pm10\ms$) 
added in quadrature to the spectrum's internal error. The 32 \kfour CCF RVs 
are presented in 
\tl{newccdccf}. The mean error, not including the rather isolated first pair 
in October 2006 (JD\,245\,4024), is $16\pm3\ms$. The substantially larger 
errors on the two RVs from 2006 are presumably related to their more distant 
acquisition time and the changing of the CCDs during 2006/7.

\bte
\bc
\caption{An abbreviated list of the relative velocity data of 32 \kfour 
detector observations acquired 2006-2013 using CCF fitting. Columns as 
for \tl{newccdiod}. The table is reproduced in full online.}
\setlength{\tabcolsep}{4pt}
\btr{crcrc}
\hline
\ml{1}{c}{JD}           & \ml{1}{c}{\vccf}   & \ml{1}{c}{$\sigma\di$} & \ml{1}{c}{bary.corr.}  & \ml{1}{c}{$S/N$}      \\
\ml{1}{c}{$245\ldots$}  & \ml{1}{c}{(\skms)}  & \ml{1}{c}{(\sms)}      & \ml{1}{c}{(\skms)}     & \ml{1}{c}{($n=110$)} \\
\hline
   4024.1070 & $   9.562 $ &     52 & $ -15.9816 $ &      123   \\
   4024.1202 & $   9.557 $ &     52 & $ -15.9813 $ &      127   \\
   5728.2797 & $  -0.638 $ &     15 & $   6.7807 $ &      186   \\
   5816.9981 & $   1.626 $ &     12 & $ -13.7272 $ &      224   \\
   5819.9725 & $   1.704 $ &     16 & $ -14.1085 $ &      117   \\
\hline
\etr
\lab{newccdccf}
\ec
\ete

It is relatively simple to check if these CCF RVs are likely to be reliable, 
since we can assess their accuracy relative to \iod RVs (\fl{ccfresids}). 
Whilst understandably of lower precision, these nineteen CCF RVs have similar 
relative accuracy as their neighbouring \iod RVs. Whilst these results 
are limited to a single star and this small sample, they suggest that 
if the weighting of the CCF orders is determined accurately, the resulting RVs 
can be of comparable accuracy to ones derived with an \iod cell. Assessing the 
credibility of the controversial \nuo planet and any competing scenarios 
depends on RVs of the highest quality. Hence we will now apply this 
order-weighting strategy to the older \kone spectra reported in \drb.

\bfi
\rotatebox{-90}{\scalebox{0.32}{\includegraphics{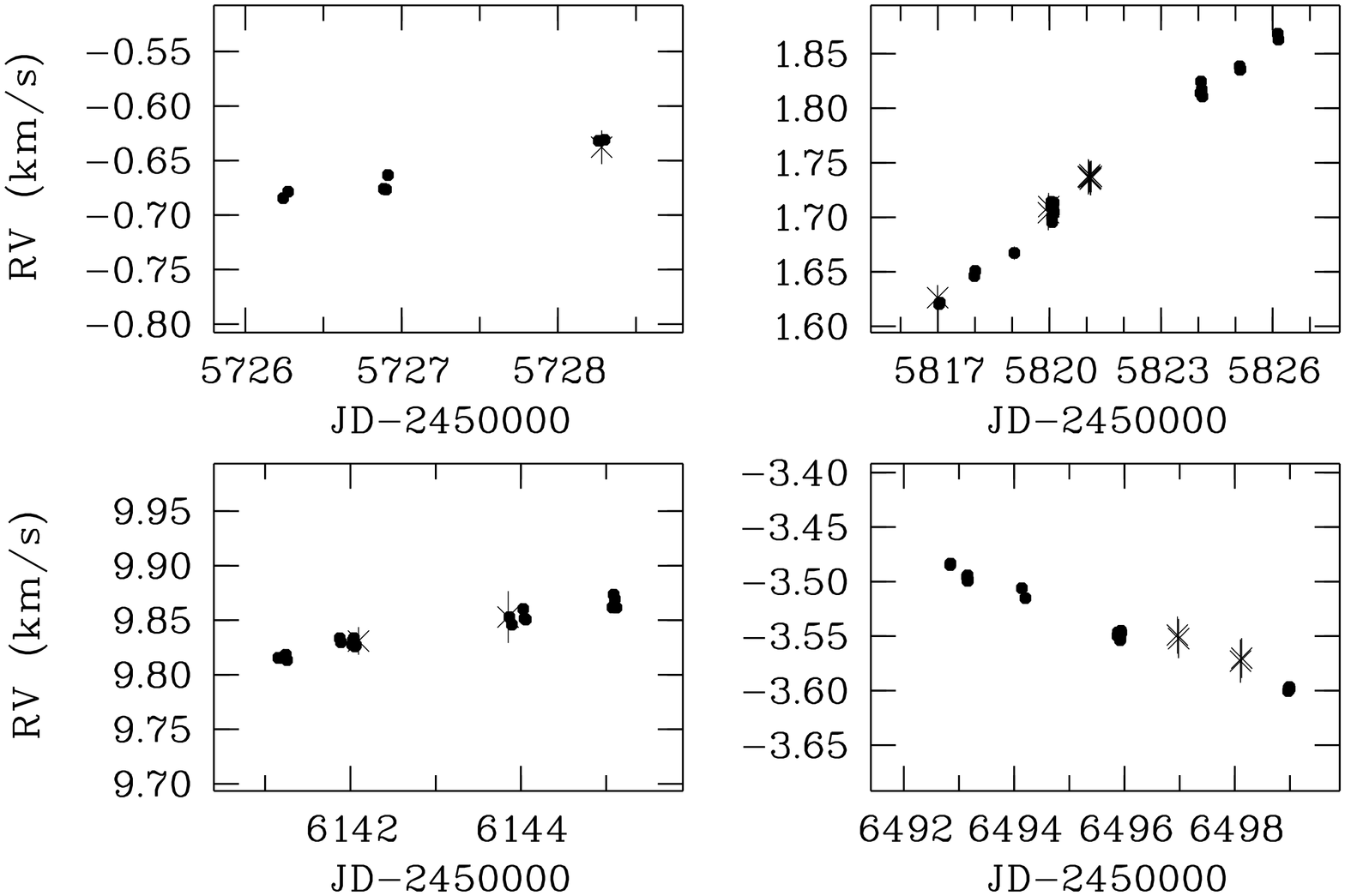}}}
\caption{Nineteen CCF RVs ($\times$) compared to all their 
neighbouring \iod RVs ($\bullet$) for the \kfour detector. The vertical 
axis has a constant span of 300\ms and any visible vertical lines are the 
corresponding 1-$\sigma$ limits.}
\label{ccfresids}
\efi

\su{Revision of \kone CCF RVs}
\lab{fixk1}
Unfortunately, the CCF RVs reported in \drb have several suspicious features 
that suggest a revision of them is prudent. For instance, the old 
internal errors (typically 4-6\ms) are similar to our \iod-RV errors, a 
Keplerian model was used to derive the spectral order weights, and perhaps 
most suspicious of all, $\sqrchi=4.2$ (\tl{orbital}), easily accounted for by 
the likelihood the errors are under-estimated.

To derive the CCF RVs from the 
561 \kfour spectra, we now have to restrict the number of examined orders 
($n=114-124$) and the width of each order ($\sim20$~\AA) to match the 
older \kone spectra. Since the 
gauge and star are the same, and the only instrumental difference is the pixel 
size, which is identical across all orders for each detector, we claim the 
weighting strategy remains valid. The simple Keplerian solutions that 
duplicate the orbital fitting used in Ramm \oo will support this claim by 
nearly halving the solution RMS.

A different reference spectrum to that used in Ramm \oo was used here, having 
a higher $S/N\sim260$, which promised somewhat better RV 
precision. To avoid the reference spectrum's internal error being zero (a 
consequence of the auto-correlation CCF process), all \kone spectra had a 
minimal jitter term of 3\ms added in quadrature to their errors, this 
being approximately the average error for the \iod RVs. This adjustment will 
be given careful reappraisal during the MCMC dynamical orbit-fitting analysis 
to give a properly refined estimate of any jitter terms. 
 
The internal errors have a number of characteristics that suggest significant 
improvements to the original ones (\fl{comperrs}): 1.~almost all are 
significantly greater than our \iod-RV errors (by a mean factor 
of about $3\times$, similar to the 2009 paper's suspicious 
\sqrchi), 2.~their mean magnitude ($15\pm4\ms$) is more 
consistent with the anticipated possible precision of a \hc-CCF orbital 
solution, and 3.~their values cover a much wider range, a 
sensible reflection of the varied observatory conditions experienced by DJR 
during 2001-2006. A plot of all our 1437 velocities is given in 
\fl{allrvs}.

\bte
\bc
\caption{An abbreviated list of the 225 spectra acquired during 2001-2007 
using the \kone CCD. Columns as for \tl{newccdiod}. The entire table is 
provided online.}
\setlength{\tabcolsep}{4pt}
\btr{crcrc}
\hline
\ml{1}{c}{JD}           & \ml{1}{c}{\vccf}   & \ml{1}{c}{$\sigma\di$} & \ml{1}{c}{bary.corr.}  & \ml{1}{c}{$S/N$}      \\
\ml{1}{c}{$245\ldots$}  & \ml{1}{c}{(\skms)}  & \ml{1}{c}{(\sms)}      & \ml{1}{c}{(\skms)}     & \ml{1}{c}{($n=110$)} \\
\hline
   2068.0607 & $  -0.346 $ &     34 & $   8.6469 $ &      128   \\
   2102.1137 & $  -2.047 $ &     19 & $   0.0825 $ &      129   \\
   2121.0920 & $  -3.256 $ &     13 & $  -4.8861 $ &      137   \\
   2180.8839 & $  -7.347 $ &     27 & $ -15.3966 $ &      134   \\
   2219.8985 & $  -9.581 $ &     18 & $ -14.5397 $ &       96   \\
\hline
\etr
\lab{oldccf}
\ec
\ete

\bfi
\bc
\rotatebox{-90}{\scalebox{0.28}{\includegraphics{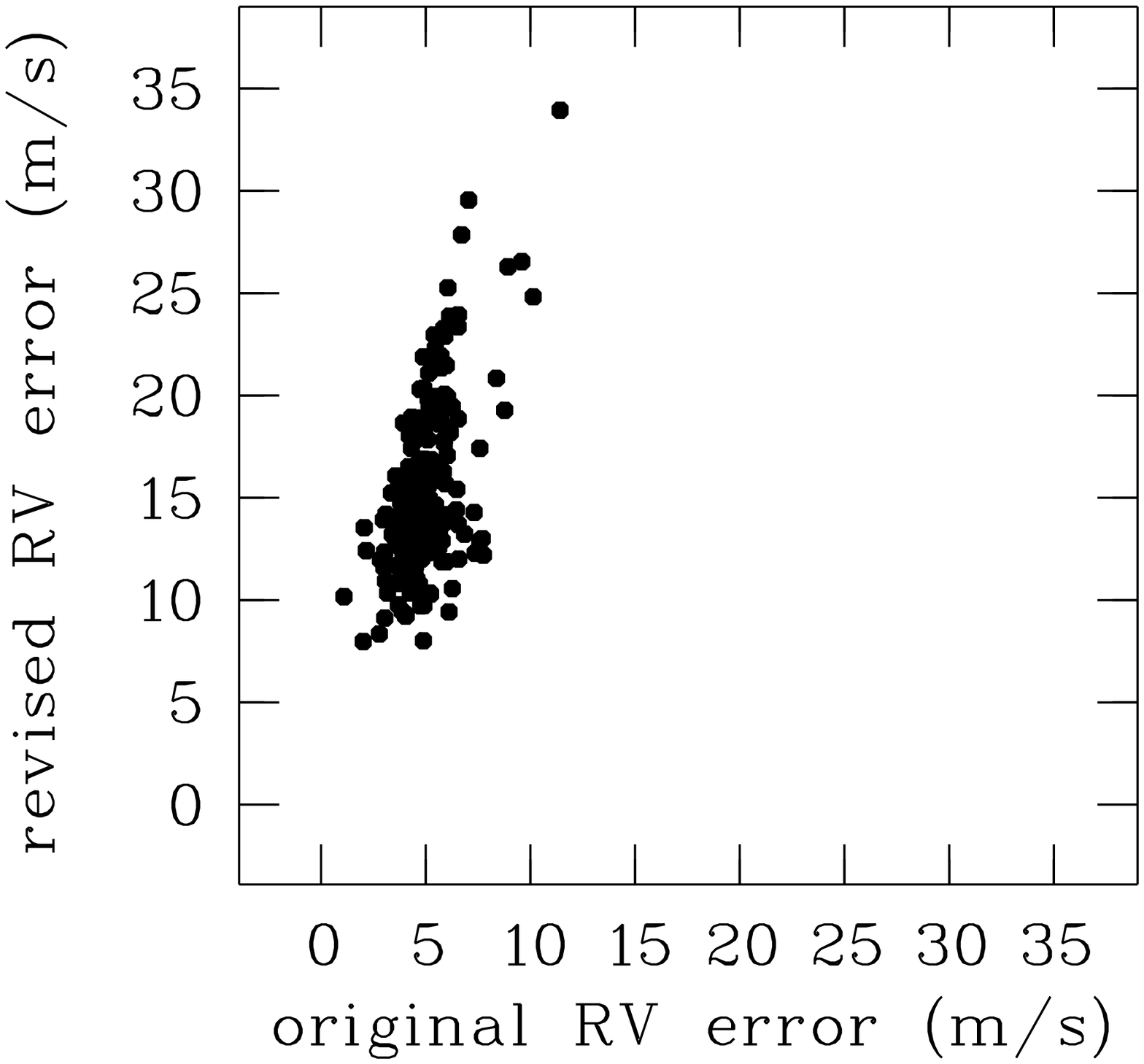}}}
\caption{A comparison of the CCF RV errors for the 204 \kone detector spectra 
common to this work and Ramm \oo (2009).}
\label{comperrs}
\ec
\efi

\bfi
\bc
\rotatebox{-90}{\scalebox{0.3}{\includegraphics{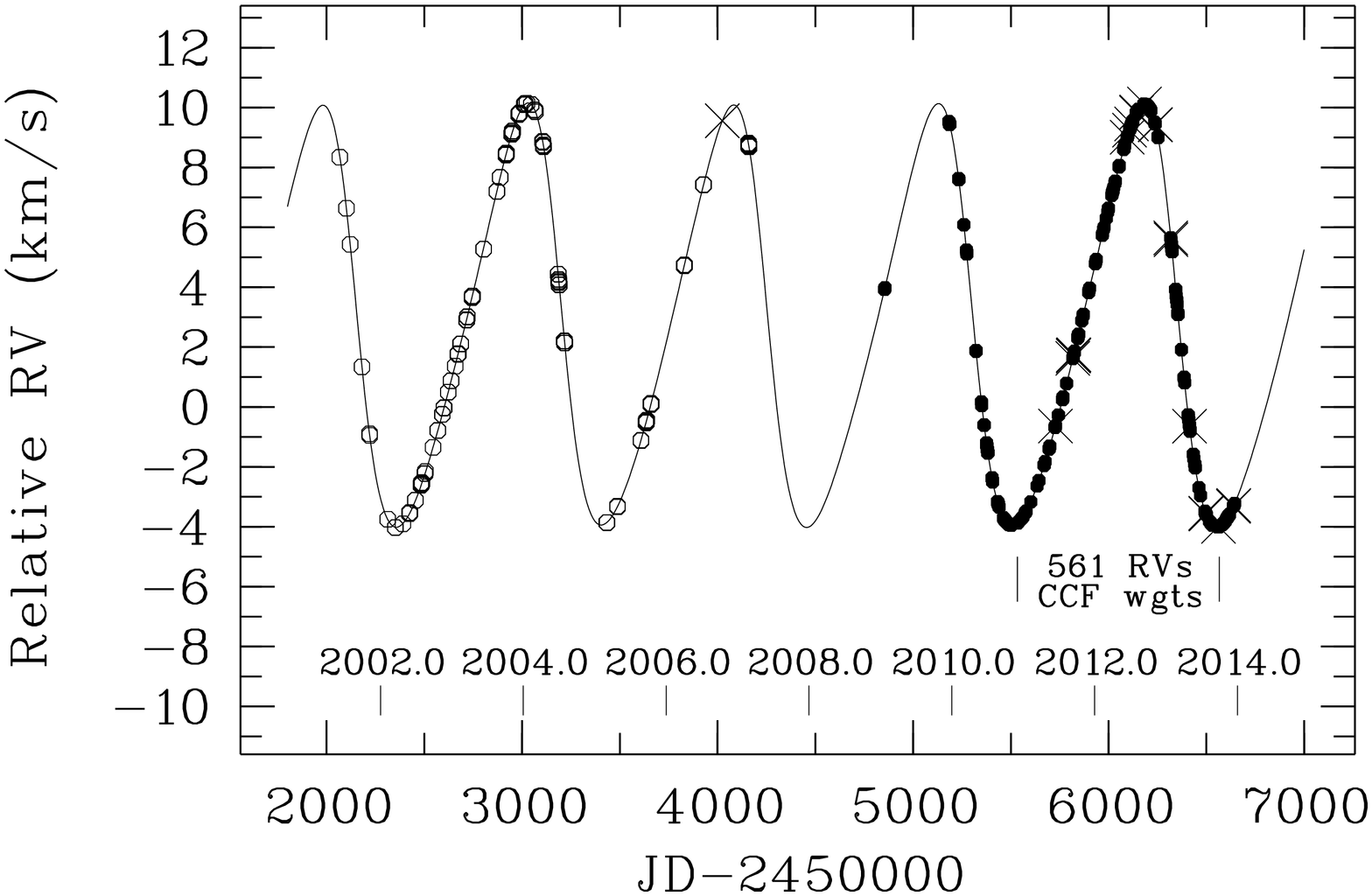}}}
\caption{The complete set of 1437 \nuo velocities 2001-2013: CCF RVs from the 
\kone detector `$\circ$'; \iod RVs from the \kfour detector `$\bullet$', CCF 
RVs from the \kfour detector `$\times$' CCF RVs. The RVs are offset to a 
common zero-point for clarity of the illustration. The small overlap of RVs 
from the two detectors is evident near JD\,4100. The series of 561 RVs used to 
re-evaluate the \kone RVs is identified (see \scl{fixk1}).}
\label{allrvs}
\ec
\efi

\su{Preliminary orbital solutions -- Keplerian}
\lab{orbfit}
Keplerian orbital solutions are not adequate to model the conjectured \nuo 
system. However they serve a useful purpose if some of their details are 
given immediately, for instance, to demonstrate the improved precision of the 
older CCF RVs (the goal of \scl{fixk1}), and to illustrate the strong 
persistent 
RV signal that all competing scenarios must explain. The actual elements are 
not critical for the paper's progress except to point out they are in close 
agreement with the values reported by \drb, in particular the elements of the 
conjectured planet include $P\dpl\sim 415$~days, $e\dpl\sim0.03$ and 
$K\dpl\sim 45\ms$. A Newtonian analysis of our RVs more 
appropriate to the conjectured strongly interacting system will be presented 
in \scl{dynsols}. The Keplerian solutions were again 
derived using a time of zero-mean longitude \tz as the reference epoch (Sterne 
1941), where the mean longitude $L=\omega + M,$ and $M$ is the mean anomaly. 

\suu{\kone detector}
Our re-evaluation provides a 
significant improvement to the precision of the 222 spectra reported in Ramm 
\oo (see \tl{orbital}): the RMS lowers to 14.4\ms and $\sqrchi=1.02$. For the 
revised set of 225 RVs (having rejected the 18 suspect 
observations described in \scl{rejects}), the RMS lowers further to 10.6\ms 
(even though the dataset's time-span is extended by 233 days by the additional 
21 spectra obtained in 2007). These results give further justification for the 
rejected spectra even though they were never tested for outlier status. A plot 
of the corresponding RV curve and residuals is given in \fl{keplcurve}. This 
is the lowest orbital RMS derived for CCF RVs obtained with HERCULES, which 
previously was limited to about 15\ms for sharp-lined spectra (see \eg 
Skuljan, Ramm \& Hearnshaw 2004; Ramm 2008), and for the first time approaches 
the spectrograph's design specification for CCF RVs ($10\ms$; Hearnshaw \oo 
2002).

\bfi
\bc
\rotatebox{-90}{\scalebox{0.3}{\includegraphics{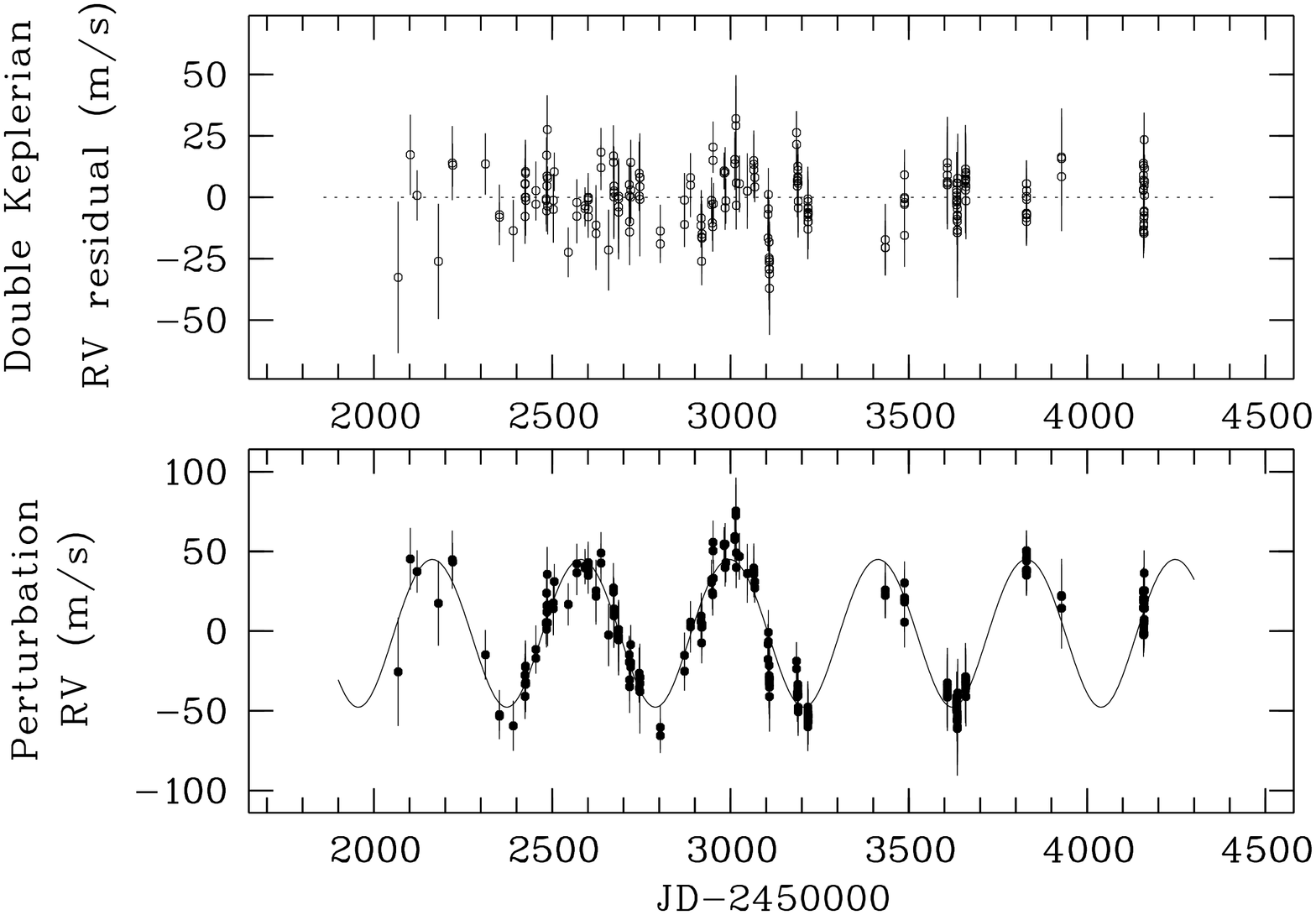}}}
\caption{The conjectured planet's radial velocities determined using a 
double-Keplerian model including the RV residuals, for 225 \kone detector 
spectra (2001--2007). The error bars represent 1\si limits.}
\label{keplcurve}
\ec
\efi

\bfi
\bc
\rotatebox{0}{\scalebox{0.4}{\includegraphics{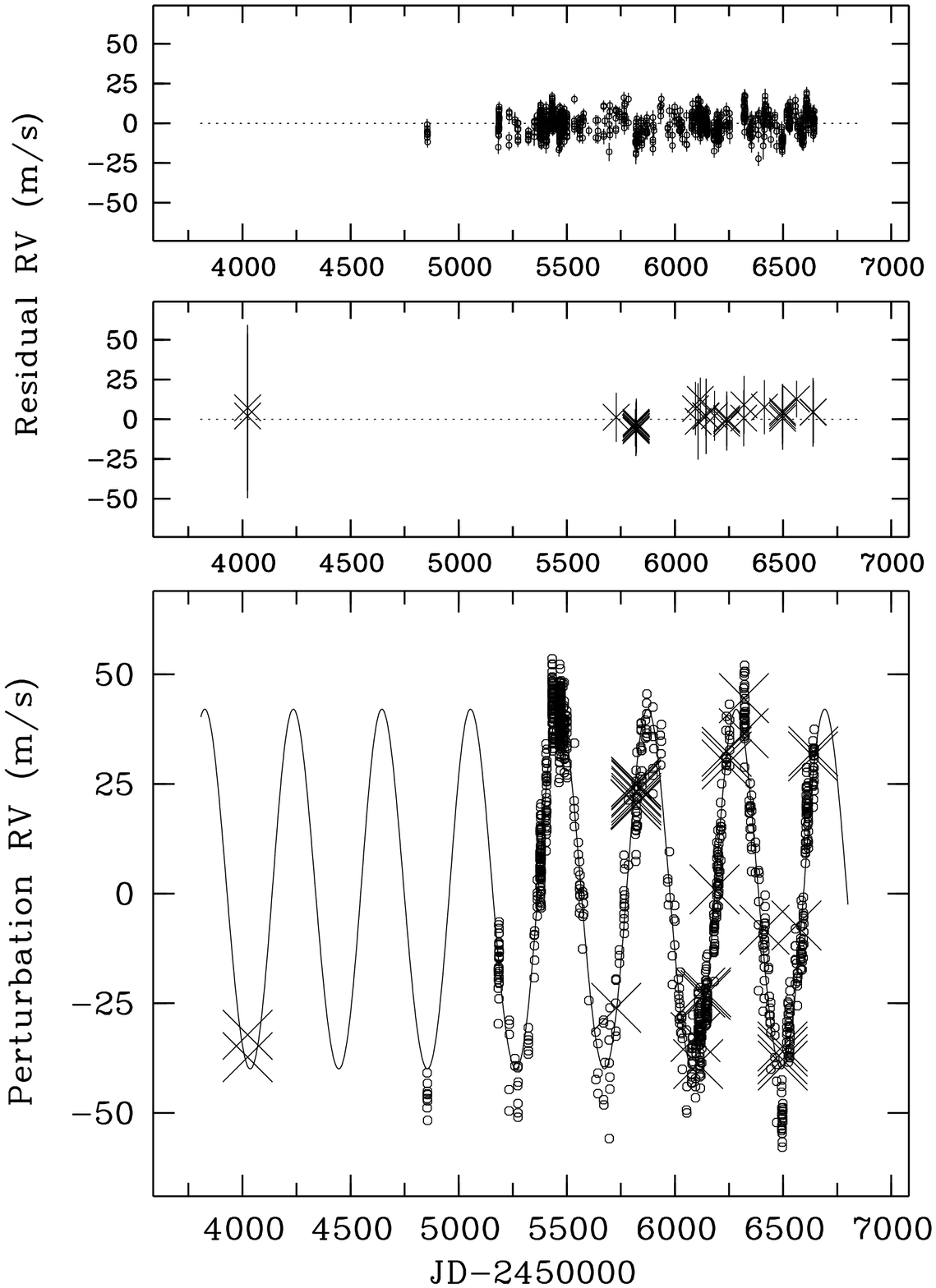}}}
\caption{The conjectured planet's double-Keplerian-model radial velocities 
including RV residuals, 1212 \kfour detector spectra (2006--2013). The 1180 
\iod-cell RVs are identified with `$\circ$'. The 32 CCF RVs are labelled with 
`$\times$'. The error bars represent 1\si limits.}
\label{iodkeplcurve}
\ec
\efi

\suu{\kfour detector}
A plot of the conjectured planet's RVs, the 
solution's best-fitting curve and the velocity residuals are given in 
\fl{iodkeplcurve}. Besides confirming the existence of the 
originally reported $\sim415$~day RV signal, the residuals of 
the 32 \iod-calibrated CCF RVs have relative accuracies that are uniformly 
similar to those solely derived from the iodine lines and AUSTRAL. 
Furthermore, the earliest two RVs, near JD\,245\,4024, have large errors and 
make no significant contribution to the orbital solution. Their high relative 
accuracy to the final fit with and without them is further 
vindication for the method used to derive these CCF RVs. The RVs and errors 
allowed a solution with an RMS of 6.5\ms and $\sqrchi=1.83$. If a further 
jitter of 5\ms is added in quadrature to the internal errors of all 1212 RVs 
the total jitter is about $6\ms$, a suitable value proposed by Johnson \oo 
(2010) for stars similar to \nuo, and $\sqrchi=1.06$ is derived.

\se{Non-RV spectral analyses}
Before describing the results of our dynamical orbital analyses we 
next present our evidence re-affirming that this periodic RV signal is 
unlikely to have a stellar origin. The investigations include those for 
chromospheric activity using the \caii H line, line-depth ratios and CCF 
bisectors, sampling respectively short, long and medium wavelengths.

\su{Ca II H\&K spectra}
The vast majority of the spectra taken with both CCDs either did not include 
the \caii H\&K orders, or did so with very inadequate $S/N$. Only one \caii 
spectrum was acquired with the \kone detector owing to the need to manually 
move the detector to record these lines -- time consuming and not ideal for 
obtaining precise RVs. Negligible chromospheric activity has been reported by 
Warner (1969) for spectra obtained during 1964-1965, whose eye-estimate value 
was `1' (the scale was 0--8 with 0 indicating absence of K-line emission).

With the \kfour CCD, RV exposures 
were far too short to provide adequate $S/N$ for the \caii lines. Therefore 
seventeen longer exposures were taken at irregular intervals approximately 
consistent with the conjectured planet's RV cycle to monitor potential 
variability. Whilst our sample is relatively small, neither the H nor the K 
line shows 
any significant variability or any periodicity consistent with the RV 
behaviour. The K~line was less well exposed and so we restrict our attention 
to the H~line.

We followed the general method of Santos \oo (2000). 
We measured the total flux of a 1~\AA\ wide window centered on the \caii H 
line ($\lambda_{\rm vac}=3969.59$~\AA ) and divided this by the total flux in 
a second region 15~\AA\ wide centered at 3990~\AA. An overlay of the 
seventeen spectra in the vicinity of the \caii H line is given in \fl{hline}, 
after Doppler shifting has been taken into account.\fn{These plots mimic the 
single spectrum reported in \drb for JD\,245\,3831 (April 2006).} The $S/N$ 
estimated at the centre of the order including the H line varied 
considerably, having an average $<S/N> = 96\pm25$. There was no correlation 
between $S/N$ and the chromospheric-activity index $S_{\rm \sss HERC}$. The 
error $\vep$ for each ratio was estimated at 10 percent\fn{Santos \oo were 
using $R=50\,000$ spectra and made this choice for one of their stars that had 
only one observation.} for $S/N=100$, and then scaled accordingly: hence 
$\vep_{(\rm S_{\rm HERC})}=10\times S_{\rm \sss HERC}/(S/N)$. The 
distribution of H-line ratios is given in \fl{hratios}. A Lomb-Scargle 
periodogram analysis was conducted and no periodicities significantly 
exceeding the background noise were found at any frequency, in particular at 
the RV-perturbation period $\sim415$~days, nor at the estimated rotation 
period of \nuo~A, $\prot\simeq140\pm35$~days (based on its radius, 
$v\sin i$ and making the simple but perhaps inaccurate assumption 
$i_{\rm \sss rot}=\ibin$).

\bfi
\bc
\rotatebox{-90}{\scalebox{0.28}{\includegraphics{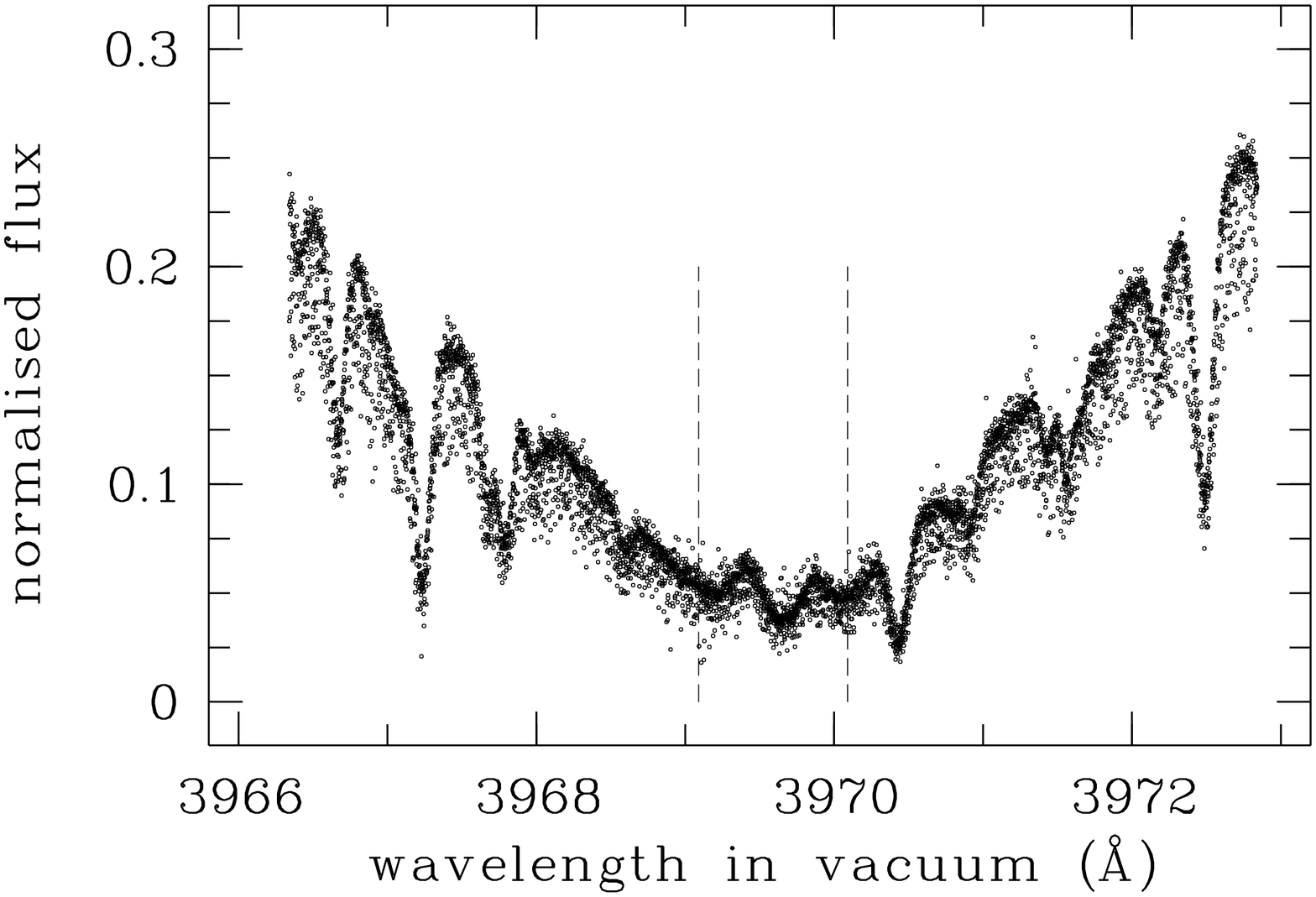}}}
\caption{The central region of the Ca II H line for seventeen \hc spectra 
recorded by the \kfour detector from 12 Sept 2011 to 18 December 2013. 
The 1~\AA\ window for measuring the total flux is identified.}
\label{hline}
\ec
\efi

\bfi
\bc
\rotatebox{-90}{\scalebox{0.28}{\includegraphics{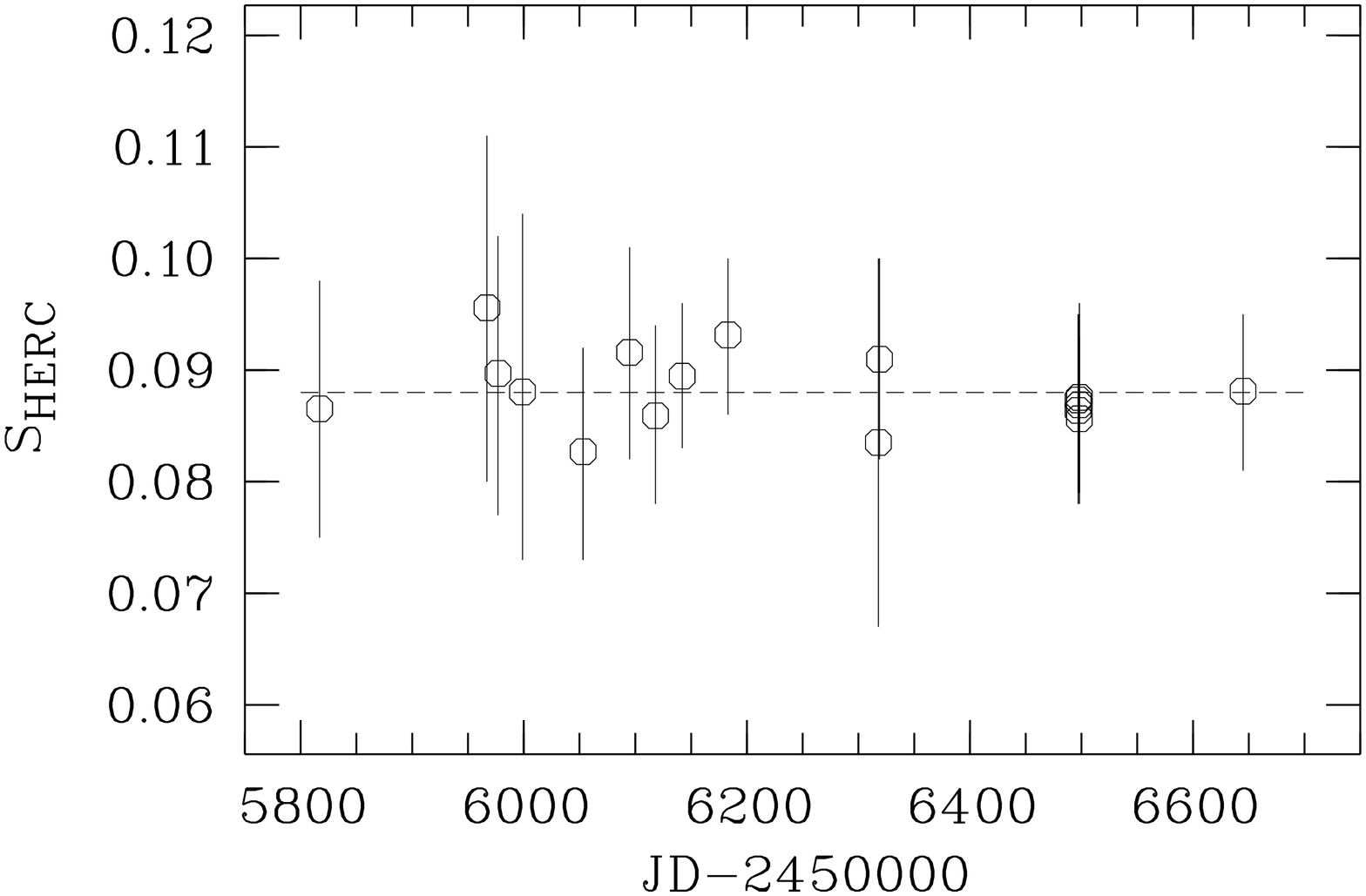}}}
\caption{The Ca II H line ratios $S_{\rm \sss HERC}$ of the seventeen 
\hc spectra illustrated in \fl{hline}. The vertical line through each ratio 
represents $\pm1\vep$.}
\label{hratios}
\ec
\efi

\su{Line-depth ratios}
\lab{ldrs}
Line-depth ratios (LDRs) have a long but infrequent history for supporting or 
discrediting exoplanet discoveries \eg 51~Peg (see \eg Gray 1997; Hatzes, 
Cochran \& Bakker 1998). In a recent paper Ramm (2015) likewise 
investigated the 215 \nuo spectra reported in Ramm \oo 
(2009), derivng 22 LDRs from ten spectral lines in the wavelength region 
6230--6260~\AA. Spectra of 20 similarly evolved stars were also 
used to create a temperature scale. Ramm recovered the original 21 
stellar temperatures with an accuracy and standard deviation of $45\pm25$~K 
and found no significant periodic behaviour in any of the LDRs.

Carefully constructed ratios are known to be extremely sensitive to 
temperature (see \eg Gray \& Johanson 1991; Gray \& Brown 2001; Kovtyukh \oo 
2003). Besides having no periodic behaviour, the 215 LDR-calibrated 
tempertaures $T_{\rm ratio}$ had a standard deviation of only 4.2~K over the 
several years 2001--2007. Making the realistic assumption \nuo~A was therefore 
not likely to be pulsating, the temperatures were converted to magnitude 
differences $\Delta m$ using the Stefan-Boltzmann law. The distribution of 
these have a striking similarity to the scatter recorded 
in the \hip photometry obtained about 15~years previously. We now extend this 
LDR analysis to the newer spectra.

Unfortunately all of the LDR spectral lines utilised in Ramm (2015) occur in 
order $n=91$, which, for the vast majority of our \kfour spectra are 
swamped with iodine lines. However, 47 
spectra were acquired without the iodine cell (some accidentally, some 
intentionally) on 21 nights during twelve observing runs spanning 2538 days 
(Oct 2006--Sept 2013), a useful additional sample. In order $n=91$, 
thirty-nine spectra have a mean $S/N=246\pm62$ and for the remaining eight it 
is $525\pm46$, the latter being part of the set of longer exposures for 
acquiring our \caii spectra. Once again, from a Lomb-Scargle analysis, there 
is no periodic behaviour in this 
latest sample in the vicinity of the $\sim415$~day RV signal, none 
consistently across all of the 22 ratios, and none with power 
significantly greater than any periodogram's background noise.

Converting each ratio into a temperature using calibration stars allows the 
temperature to be derived more precisely by the averaging of all the results 
for each spectrum. These in turn can be converted to magnitude 
differences that can further act as a guide for assessing the reliability of 
the method. The average temperature for all 22 ratios is $4811\pm28$~K. This 
is consistent with the effective temperature reported by Fuhrmann \& Chini 
(2012; $4860\pm40$~K). The 47 averaged temperatures have a mean of 
$T_{\rm \nu Oct}=4811\pm4.1$~K, practically identical to the value reported in 
Ramm (2015; $4811\pm4.2$~K).

\bfi
\bc
\rotatebox{-90}{\scalebox{0.3}{\includegraphics{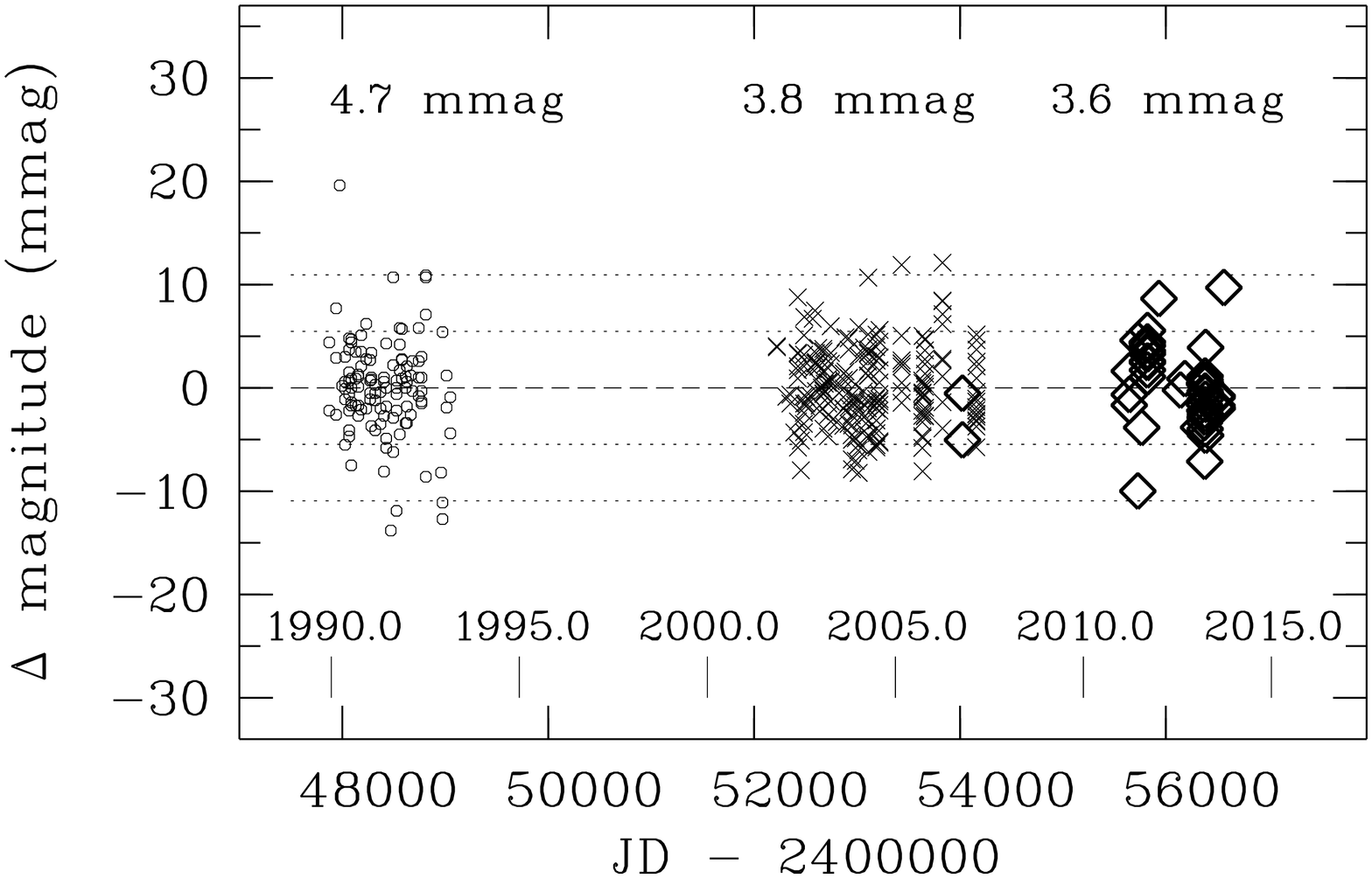}}}
\caption{The magnitude differences, $\Delta m$, for \nuo~A based on 
22 line-depth-ratio calibrated temperatures: 215 \kone spectra: `$\times$', 
and 47 \kfour spectra: `$\lozenge$'. All spectra acquired with resolving power 
$R=70,000$. These are compared to 116 best-quality \hip observations 
($\rm flag = 0$: `$\circ$'). The dotted lines identify the $\pm1.5\si$ and 
$\pm3\si$ limits from the \kone predictions. The value above each dataset is 
the $\Delta m$ standard deviation.}
\label{mmags}
\ec
\efi

Magnitude-differences were derived and these again have a striking resemblance 
to the brightness variations recorded by \hip and estimated by Ramm (2015), 
the three sets of $\Delta m$ values now spanning about 23 years (\fl{mmags}). 
It should be noted that our estimates for $\Delta m$ are 
extremely sensitive to the temperature ratio $T_{\rm ratio}/T_{\nu \rm Oct}$: 
varying $T_{\rm ratio}$ by only one degree changes 
$\Delta m$ by about 1~mmag. Such similar behaviour of three high-precision 
sets of $\Delta m$ values would seem to be unlikely to be due to misguided 
analytical methods. It would also be surprising if, whilst much 
larger variations did exist, they only occurred in the years not sampled. 

\su{CCF bisectors}
Spectral line 
bisectors (see \eg Gray 1983; Hatzes \oo 1998; Povich \oo 2001; Gray 2005) and 
that of CCFs (Queloz \oo 2001; Dall \oo 
2006; Ba\c{s}t{\"u}rk \oo 2011) are both well-recognised means to assess if a 
star's surface dynamics and irregular features can significantly influence 
RVs, or to identify spectral contamination from previously unrecognised 
stellar companions (Wright \oo 2013). Bisectors have also been shown to vary 
in systematic ways with surface gravity (\ie luminosity) and effective 
temperature (Gray 2005; Gray 2010; Ba\c{s}t{\"u}rk 
\oo 2011) though without the extreme sensitivity that for instance line-depth 
ratios have for temperature.

Our bisectors were derived from 675 \kfour spectra (acquired Dec 2010 -- Dec 
2013) for the \iod-free orders with higher $S/N$, $n=114-120$. The CCFs were 
created using a template spectrum with $S/N\sim240$.  Five CCF flux 
levels were measured, at 20, 30, 45, 60 and 75\% of the CCF's parabola-fitted 
peak value. We constructed two variations of commonly used bisectors, the 
velocity span, $V_{\rm sp}=v_{30}-v_{75}$, and the bisector inverse slope 
$BIS=(v_{20}+v_{30})/2 - (v_{60}+v_{75})/2$. We found no evidence of any 
significant periodic behaviour for any order corresponding to the RV signal at 
$\sim415$~days from our Lomb-Scargle analyses, just as was reported in the 
previous CCF bisector analysis of \drb. There is no correlation between the 
bisector values and the spectra's $S/N$.

To illustrate our bisector results we concentrate on the BIS results. To 
provide them in in a manner we believe is more revealing, we divided the 
spectra into sets defined to be separate `observing runs'. Each `run' only 
included spectra with no more than 3.5 days between consecutive dates. 
Typical observing runs lasted about a week and were separated by 2--4 weeks, 
but occasionally circumstances interrupted longer runs by several days (\eg as 
a result of instrument failures or adverse weather), effectively separating 
groups of spectra. In most cases the subsets of spectra represent true 
separate observing runs. About half of these subsets included eight or more 
spectra, and the maximum number of nights included in a single `observing run' 
was fourteen (two runs).

\bfi
\bc
\rotatebox{-90}{\scalebox{0.33}{\includegraphics{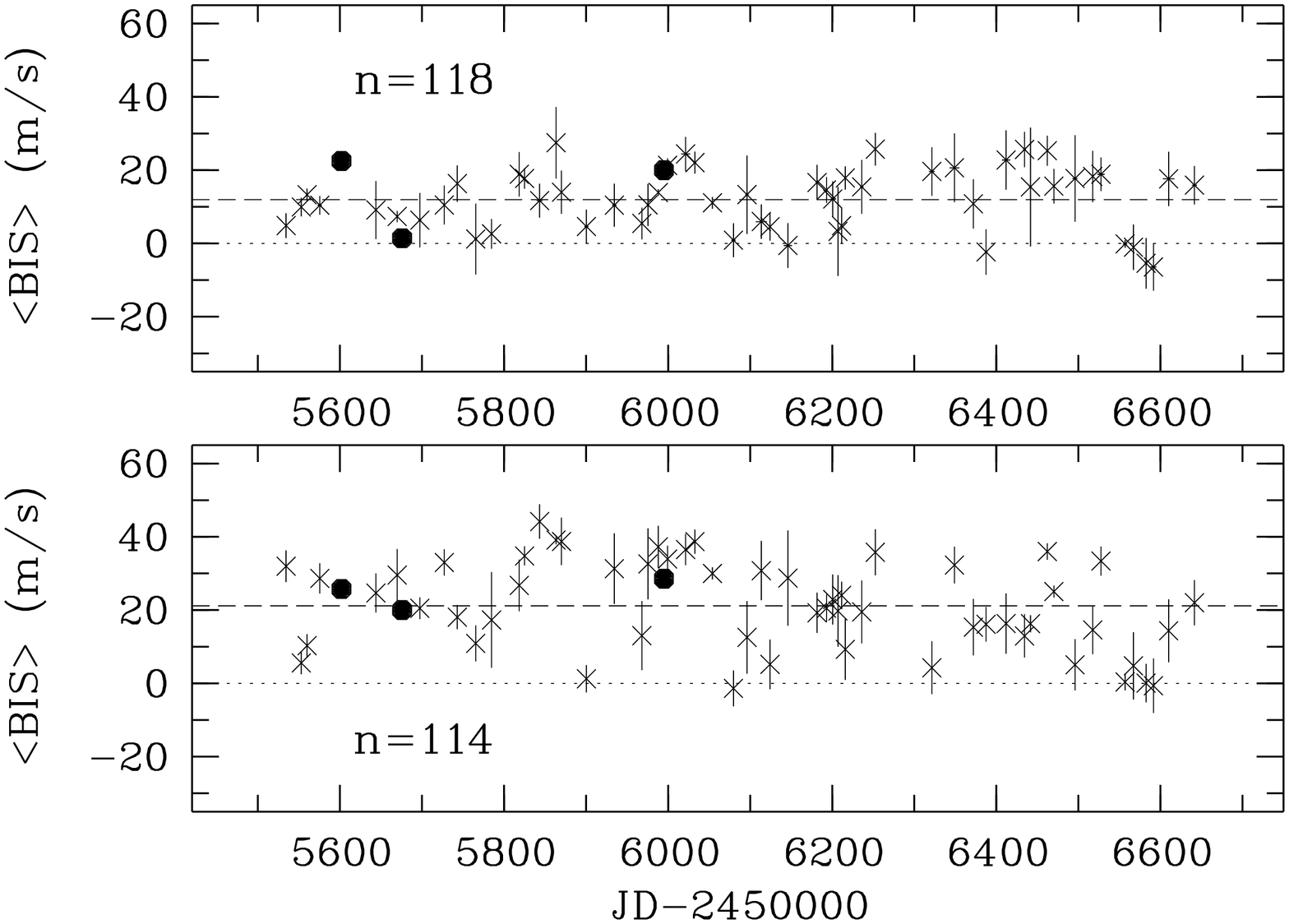}}}
\caption{The mean bisector inverse slopes ($\langle BIS\rangle$) for 59 
`observing 
runs' as defined in the text, derived from the CCFs of 675 spectra. Orders 
$n=114$ and $n=118$ are shown, together with their 1\si standard deviations. 
`Observing runs' with only one BIS value are labelled with `$\bullet$'. The 
average of each order's mean is represented by a dashed line.}
\label{bisectors}
\ec
\efi

We derived the mean BIS 
$\langle BIS\rangle$ and standard deviation for the 59 allocated `observing 
runs'. We discuss the results for $n=114$ and $n=118$ as these have the 
least variability (\fl{bisectors}). These standard deviations, 
$\sigma_{\rm \langle BIS\rangle}$, for many subsets is very low, and 
averages $5\pm3\ms$, or about 5\% of the total BIS range for that 
order.\fn{This low fraction, a few percent, is true for all seven orders.} For 
both orders, only two spectra were rejected (their 
BIS values were treated as outliers as they each were about $4\sigma$ from the 
mean). For the ten runs that included 20 or more spectra (to a maximum of 67 
spectra), $\sigma_{\rm \langle BIS\rangle}\sim7\pm2\ms$, and for the two runs 
that extended over fourteen nights $\sigma_{\rm \langle BIS\rangle}\sim8\ms$. 
The $\langle BIS\rangle$ values for these two orders have ranges significantly 
less than the RV signal's amplitude and are only weakly correlated ($r=0.36$), 
whereas the 675 RVs are perfectly correlated ($r=1.000$). Thus we claim that 
most of the scatter is not due to measurement errors, 
otherwise, firstly, the scatter would presumably be evident in the 
individual runs. Since our spectra can easily reveal the signal in our CCF RVs 
($K\dpl\sim40-45\ms$), we also claim these very small `observing run' standard 
deviations imply we should be able to easily detect the perturbation signal in 
our CCF bisectors if it was present. The distribution of BIS mean values seems 
more likely to be due to acquisition variations, a result of the reduction 
process and/or stellar variability unrelated to the RV signal. 

\se{Dynamical orbital solutions}
\lab{dynsols}
We derived dynamical orbital solutions for the 225 \kone detector RVs and 
the full set of 1437 RVs, the latter being the global solution we desire for 
our stability modelling. We characterize the masses and orbits of the \nuo 
components using an $N$-body 
differential evolution Markov Chain Monte Carlo (\RUNDMC, Nelson \oo 2014a), 
which incorporates the \SWARMNG framework to integrate planetary systems on 
GPUs (Dindar \oo 2013). We adopt our prior probability distribution and 
likelihood function from Nelson \oo (2014a). Additional orbital parameters now 
include the 
primary star's mass \cp, each companion's mass (\cs and the conjectured planet 
$\mass\pl$), and the mean anomaly $M$ at our chosen epoch, plus the RV 
zero-point offset $\delta V$, and jitter $\sigma_{\rm jit}$, all based in a 
Jacobi coordinate system. The binary-orbital solution 
reported in \drb 
provides a precise estimate of the binary's longitude of line-of-nodes 
$\Omega\dbin$ and its inclination \ibin (\tl{orbital}) and therefore our 
reported orbital 
models fixed these two parameters.  We also performed 
a similar run for each case that did not impose astrometric constraints and 
found the parameter estimates to be quantitatively similar. \RUNDMC has been 
used to analyze other strongly interacting planetary systems while 
allowing for mutual inclinations (e.g. 55 Cancri, Nelson \oo 2014b; Gliese 
876, Nelson \oo 2016).
On one hand, \nuo has a much lower dimensional model (i.e. only three bodies 
and fewer offset/jitter parameters); on the other hand, the proximity of these 
masses suggests the presence of extremely strong gravitational interactions, 
which can result in a posterior distribution that is very difficult to sample 
from.

Considering the lessons from Nelson \oo (2014a) regarding how to explore 
parameter space efficiently, we set the following algorithmic parameters for 
\RUNDMC: $n_{\rm chains}=300$, $\sigma_\gamma=0.01$, and MassScaleFactor=1.0.
We set our integration time-step to roughly 84 minutes and use the 
time-symmetrized Hermite integrator (Kokubo et al. 1998). The epoch of these 
solutions is the time of the first observation, \ie JD245\,2068.0607 for each 
dataset. 

We sampled from the Markov chains after they had burned-in sufficiently 
and obtained a set of posterior samples. For the RVs acquired only with the 
\kone detector, summary statistics of the model parameters are given in 
\tl{dyn1k} and corresponds to $\sqrchi=0.85$. A similar table for the total 
set of RVs are given in \tl{dynall} and has $\sqrchi=0.99$.

\bte
\bc
\setlength{\tabcolsep}{4pt}
\btr{lcc}
\hline
                            &   \ml{2}{c}{\nuo A's absolute orbit} \\  
\hline
  companion                 &\hgap  \nuo~B       & conjectured planet \\[1ex]
$\mass$(\msun)        &\hgap $  0.584\,8\spm{0.0003}$              & $0.002\,5\spm{0.0003}$ \\[1ex]
  $K_1$ (\kms)        &\hgap $  7.0514\spm^{0.0047}_{0.0056}         $ & $ 0.0465\spm^{0.0063}_{0.0041} $ \\[1ex]
  $e$                 &\hgap $ 0.236\,48\spm^{0.000\,61}_{0.000\,70} $ & $ 0.116\spm^{0.078}_{0.067}    $ \\[1ex]
  \omp (\dg)          &\hgap $ 75.01\spm^{0.15}_{0.16}               $ & $ 94\spm^{55}_{227}            $ \\[1ex]
  $P$ (days)          &\hgap $ 1050.92\spm^{0.46}_{2.18}             $ & $ 418\spm^{19}_{17}            $ \\[1ex]
  $a$ (au)$\dag$      &\hgap $  2.630\,00\spm^{0.000\,70}_{0.003\,61}$ & $ 1.283\spm^{0.038}_{0.036}    $ \\[1ex]
  $\Omega$ (\dg)      &\hgap    87.0 (fixed)                          & $ 252\spm^{38}_{15} $    \\[1ex]
  $i$ (\dg)           &\hgap    70.8 (fixed)                          & $ 100\spm^{19}_{46} $    \\[1ex]
  $M$ (\dg)           &\hgap $ 339.26\spm^{0.15}_{0.14}              $ & $ 183\spm^{227}_{50} $  \\[1ex]
$\Phi_{\rm Bb}$ (\dg) & \ml{2}{c}{$147\spm^{11}_{27}$}                                   \\[1ex]
 RMS (\ms)            &  \ml{2}{c}{11}             \\[1ex]
  $\chi^2$            &  \ml{2}{c}{$152.2\spm^{6.1}_{5.7}$}             \\[1ex]
  $\chi^2_{\rm eff}$  &  \ml{2}{c}{$138.4\spm^{6.1}_{5.6}$}                                 \\[1ex]
  RV offset (\ms)     &  \ml{2}{c}{$-6044.0\spm^{1.8}_{0.6}$}                                \\[1ex]
  jitter (\ms)        &  \ml{2}{c}{$0.81\spm^{1.2}_{0.6}$}                                  \\[1ex]
\hline
\etr
\ca{Dynamical solutions for \nuo~A and its conjectured planet using 
225 CCF RVs from \kone detector observations (2001-2007). Estimates are 
computed using 15.9, 50, and 84.1 percentiles. $\Phi_{\rm Bb}$ 
is the mutual inclination and $\chi^2_{\rm eff}$ is the standard $\chi^2$ 
with an additional penalty based on the jitter value similar to that described 
in Nelson et al. (2014a). $\dag$ Note: the semimajor axis is of the relative 
orbit $a=a_{\rm \sss A}+a_{\rm \sss comp}$.}
\lab{dyn1k}
\ec
\ete

\bte
\bc
\setlength{\tabcolsep}{2pt}
\btr{ccccc}
\hline
      &   &                  &   \ml{2}{c}{\nuo A's absolute orbit} \\  
\hline
\ml{3}{l}{companion}         &\ngap   \nuo~B       & conjectured planet \\[1ex]
\ml{3}{l}{$\mass$(\msun)}    &\ngap $0.585\,22\spm{0.000\,04}$ & $0.002\,01\spm{0.000\,02}$\\[1ex]
\ml{3}{l}{$K_1$ (\kms)}      &\ngap$  7.055\,44\spm^{0.000\,40}_{0.000\,44} $ & $ 0.038\,74\spm^{0.0009}_{0.0012} $\\[1ex]
\ml{3}{l}{$e$}                &\ngap $ 0.236\,80\spm{0.000\,07}               $ & $ 0.086\spm^{0.043}_{0.036} $ \\[1ex]
\ml{3}{l}{\omp (\dg)}         &\ngap $ 74.970\spm{0.016}                      $ & $ 350\spm^{8}_{25}          $ \\[1ex]
\ml{3}{l}{$P$ (days)}         &\ngap $ 1050.69\spm^{0.05}_{0.07}              $ & $ 414.8\spm^{3.6}_{2.6}     $ \\[1ex]
\ml{3}{l}{$a$ (au)}           &\ngap $  2.629\,59\spm^{0.000\,09}_{0.000\,11 }$ & $ 1.276\spm^{0.007}_{0.005} $ \\[1ex]
\ml{3}{l}{$\Omega$ (\dg)}     &\ngap    87.0 (fixed)                            & $ 237.8\spm^{0.8}_{0.6}     $  \\[1ex]
\ml{3}{l}{$i$ (\dg)}          &\ngap    70.8 (fixed)                            & $ 112.5\spm^{2.4}_{1.5}     $  \\[1ex]
\ml{3}{l}{$M$ (\dg)}          &\ngap $ 339.286\spm^{0.023}_{0.019}            $ & $ 301\spm^{22}_{7}          $ \\[1ex]
\ml{3}{l}{$\Phi_{\rm Bb}$ (\dg)} &\ml{2}{c}{$152.5\spm^{0.7}_{0.6}$}                \\[1ex]
\ml{3}{l}{RMS (\ms)}          & \ml{2}{c}{8.4}                                     \\[1ex]
\ml{3}{l}{$\chi^2$}            & \ml{2}{c}{$1396.2\spm^{53.7}_{54.4}$}               \\[1ex]
\ml{3}{l}{$\chi^2_{\rm eff}$} & \ml{2}{c}{$2969.9\spm^{7.9}_{5.5}$}                 \\[1ex]
\hline
 detector & RV method & $N$   &\wide  RV offset (\ms)                    & jitter (\ms)                \\
\hline
\kone     &  CCF      & 225   &\wide $ -6045.4\spm^{1.1}_{1.0}$         & $ 4.1\spm^{2.4}_{2.9}$      \\[1ex]
\kfour    &  CCF      & 32    &\wide $ 2644.5\spm^{2.7}_{2.9}$          & $ 1.2\spm^{2.1}_{0.8}$      \\[1ex]
\kfour    & \iod      & 1180  &\wide $ 2642.01\spm^{0.29}_{0.32}$       & $ 5.89\spm^{0.17}_{0.18}$   \\
\hline
\etr
\ca{Dynamical solutions for the \nuo~A and its conjectured planet using 
all 1437 \hc observations (2001-2013). See \tl{dyn1k} for some explanations.}
\lab{dynall}
\ec
\ete

\se{Stability modelling}
Our posterior samples were integrated for $10^6$ years using a 
Jacobi-coordinate-system variant of the general Bulirsch-Stoer (B-S) 
integrator in the $N$-body dynamics package \MERCURY (Chambers 1999). A 
Bulirsch-Stoer integrator, similarly used by 
Go{\'z}dziewski et~al. (2013) in their detailed analysis of the \nuo system, 
is appropriate given the assumed geometry includes a nearby massive 
secondary-star perturber, and which makes for instance the `hybrid' integrator 
unsuitable. Nothing remained dynamically stable past our integration 
baseline.

\su{Grid search for stable models: methods and results}
\lab{stabgrid}
Given the lack of promising results with our posterior samples, a grid search 
was also undertaken. The grid was created from the planet best-fitting planet 
elements $a$, $e$, $i$, $\omega$, and $M$, encompassing $\pm3\sigma$ across a 
uniformly divided 
range for each. Since $\Omega\pl$ was relatively precise compared to the 
other orbit angles, it was also kept constant. The grid was composed of 
$9\times15\times13\times11\times10=193,050$ models, the respective grid steps 
were $\Delta a\pl=0.00412$~au, $\Delta e\pl=0.015$, $\Delta i\pl=1\dg$, 
$\Delta \om\pl=10\dg$, $\Delta M\pl=10\dg$. The binary elements, which have 
relatively very high precision, were kept constant at their best-fitting 
values.

Given the models' exceptional geometry the choice of a reliable time-step 
$\tau$ was also given careful consideration. Ultimately it was decided a 
suitable strategy was to process the entire grid for several time-steps. 
Small but, critically, insignificant survival-time differences for each model 
would give us confidence the integrator was likely to be reliable for this 
extreme system. The shortest time-step was relatively brief, 0.4~days 
($\sim 10^{-3}P\pl$), and the three others were simply prime numbers 3, 
7 and 11~days, these chosen to be integers and non-integer multiples of 
any other. Finally, for the two boundary time-steps 0.4 and 11 days, we 
extended our integration times to $10^6$~yrs. The resulting survival times 
were indeed very similar.\fn{For instance, for the four time-steps, the mean 
number of survivors to $10^3$~yrs was $35,351\pm24$ and to $10^5$~yrs it was 
$17,334\pm52$. The survival times for the individual models were within 
5\% of each other in nearly 85\% of them.}

The time-step chosen for the grid results reported here, $\tau=11$~days, was 
therefore not at all apparently critical, but being less CPU-time demanding, 
it was the only time-step that had its maximum integration time extended 
further, for all models to 
$10^7$~years. The model did not survive if the planet 1.~travelled beyond 5 
au (about twice \abin; 89\% of the models), 2.~collided with the 
central/primary star (5\%), or 3.~collided with the secondary (0.3\%). Thus 
about 5\% of our grid models (9,786 models) survived to $10^7$ years. Most of 
the grid parameters (\ie \apl, \epl, \ipl, $M\pl$) have gently sloping 
distributions across their entire ranges for the fraction of survivors to 
$10^7$ years. The mean longitude $L=\omega + M$ distribution differed in that 
it was strikingly symmetric and narrow. This symmetry gave us further 
confidence our B-S integrator was reliable. The fraction of models surviving 
to $10^3$, $10^5$ and $10^7$ years are illustrated in \fl{meanlong}. The 
models most likely to survive, even to $10^3$ years, had initial 
$L\pl=310\pm10\dg$.

\bfi
\bc
\rotatebox{-90}{\scalebox{0.32}{\includegraphics{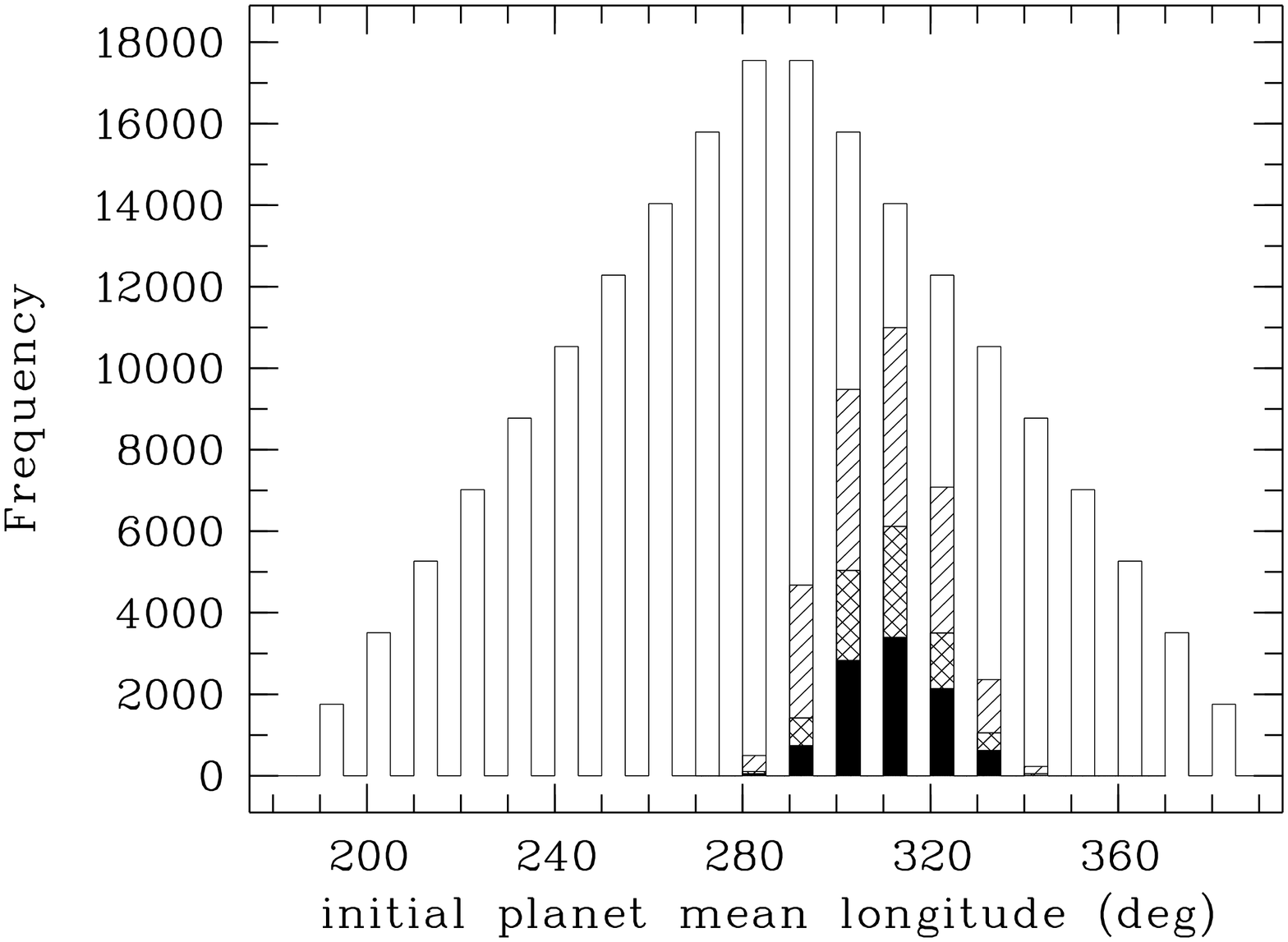}}}
\caption{Histogram of the full grid of initial mean longitudes $L$ for our 
193,050 stability models with a time-step of 11~days. The lightly hatched 
subset survived to $10^3$ years (35,327 models), the double hatched to 
$10^5$~yrs (17,281 models), and the bold to $10^7$ years (9,786 models). 
$L>360\dg$ has been plotted to more easily illustrate the distributions.}
\label{meanlong}
\ec
\efi

\bfi
\bc
\rotatebox{0}{\scalebox{0.52}{\includegraphics{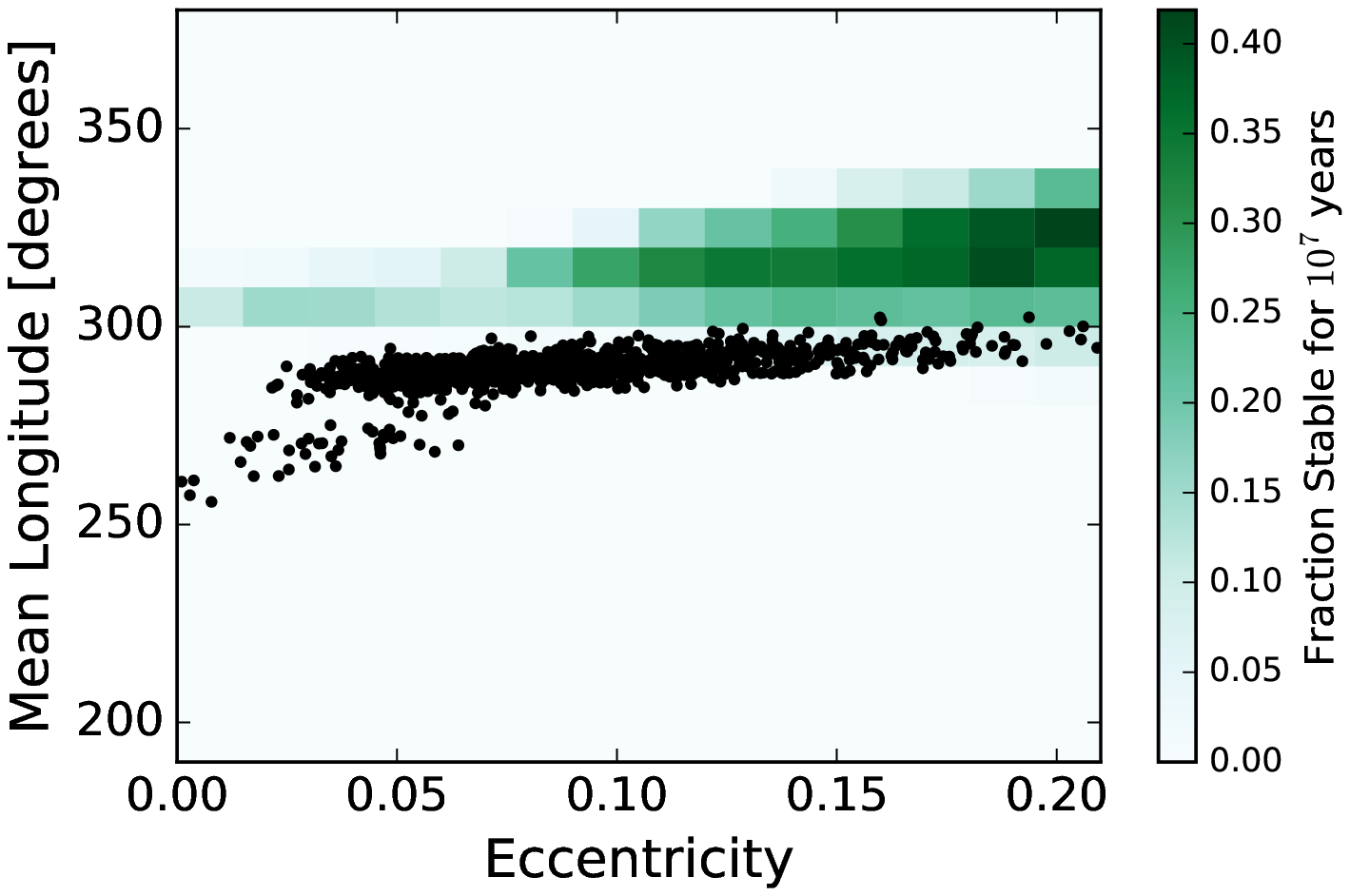}}}
\caption{Grid search results in planet mean longitude and eccentricity space. 
Darker shades represent a higher fraction of stable systems residing within 
that particular cell (color scale on right). Black dots are posterior samples. 
The online version is in colour.}
\label{stabilities}
\ec
\efi

The distribution of stable models with respect to the planet's initial 
eccentricity and mean longitude is included in \fl{stabilities}. We find 
higher eccentricities tend to provide more stable models. One model that 
survived to $10^7$ years had grid values all within 1-sigma of the 
best-fitting elements. Due to our computer-time limitations, only this one had 
the integration time extended to a little beyond $10^8$ years. One model 
survived the full integration for two time-steps, 0.4 and 11 days, but the 
planet was ejected at 35 and 91 Myrs for the 3-day and 7-day time-steps 
respectively. This model's parameter set is given in \tl{bestmod}.

\bte
\bc
\setlength{\tabcolsep}{4pt}
\btr{lcccccc}
\hline
companion & $a$ (au) & $e$ & $i$ (\dg) & $\omega$ (\dg) & $\Omega$ (\dg) & $M$ (\dg) \\
\hline
\nuo B &  2.62959  &  0.2368  &  70.8 & 75.0   &   87.0      &  339.3   \\[1ex]

planet    &  1.27261  & 0.12 &  115.0    & 355.0  &  237.8      &  315.0 \\
\hline
\etr
\ca{The single model found to survive $10^8$ years within 1-\si of our 
dynamical solution's best fit (given in \tl{dynall}). The component masses 
were $\cp=1.61\msun$, $\cs=0.58522\msun$ and $\mass\pl=0.00201\msun=2.1\mjup$.}
\lab{bestmod}
\ec
\ete

\se{Other considerations}
The controversial $\sim415$-day RV signal demands every possible 
characteristic of \nuo be considered. Here we discuss the system's space 
velocities, future imaging and astrometry, and the nature of the 
conjectured planet's discovery.

\su{Space velocities and age estimate}
\lab{age}
The conjectured \nuo planet poses a challenge with regards its formation and 
some dynamical process such as a gravitational event including a passing star 
would 
appear to be amongst the most likely. Hence it seems worthwhile to determine 
\nuo's present space velocities, although these results presumably place 
only a weak constraint on any past event. Quite reliable estimates should be 
achievable given the recent updating of all the required parameters. Space 
velocities inconsistent with the estimated age might imply such a dynamical 
process was more likely, though clearly too violent an event would completely 
disengage all the system's components.

The improved barycentric radial velocity $\gamma$, parallax $\varpi$ and 
proper motions from Ramm \oo (2009) have been used to re-assess the 
heliocentric space velocities $(U, V, W)$ and the corresponding probability 
of \nuo's membership of the thin or thick disc as an age 
indicator.\fn{$\gamma = +35.24\pm0.02$\kms; $\varpi=45.25\pm0.25$~mas; 
$\mu_\alpha\cos\delta = +52.58\pm0.53$~mas/yr; 
$\mu_\delta = -240.80\pm0.47$~mas/yr: note the correction to the sign of 
$\mu_\delta$ given in Ramm \oo.} The 
calculation was performed using the procedure described in Bensby \oo (2003) 
together with the equatorial coordinates (ESA 1997), yielding 
$U=+7.53\pm0.16\kms$, $V=-41.43\pm0.26\kms$ and 
$W=-11.62\pm0.11\kms$. These velocities were then converted to the local 
standard of rest (LSR): $U_{\sss \rm LSR}=+17.5\kms$, 
$V_{\sss \rm LSR}=-36.2\kms$ and $W_{\sss \rm LSR}=-4.5\kms$ (see Dehnen \& 
Binney 1998). These values in fact differ insignificantly from those 
reported by Bartkevi\u{c}ius \& Gudas (2002). When plotted on a Toomre 
diagram, \nuo is placed on the outer edge of the intermediate-age thin-disc 
population (age $\sim$~3-4 Gyr), in close agreement with that given in 
\tl{stellar}.

\su{Future astrometry and imaging}
Any dynamical orbital solutions, stability studies and efforts to understand 
the possible formation 
scenarios will also be able to be conducted with greater confidence if \nuo's 
stellar masses and astrometry are more accurately determined. Ideally, this 
should be done in an as model-free manner as possible. Direct imaging and 
astrometric observations would also help define the true nature of \nuo~B 
since the contrast observed would determine if it was a late-type 
main-sequence star or a white dwarf (WD). Assuming the stellar components are 
coeval, if \nuo~B is found to be a WD, which presumably could also reveal 
itself by a UV excess, it would have had to evolve much 
faster than the present primary, shedding at least one solar mass in the 
process (the mass difference of the present stellar components).

Since an astrometric orbit for the binary of reasonable 
quality has now been determined, in principle, the stellar masses can be 
derived by a single astrometric observation (the angular separation $\rho$ and 
the position angle $\theta$). One such 
interferometric observation was reported several decades ago. Unfortunately, 
whilst Morgan, Beckmann \& Scaddan (1980) claimed success with their single 
observation ($\rho=0\upp104$, $\theta=331\dg$), and their separation is 
approximately consistent with our prediction ($\rho\sim0\upp12$), 
their position angle is not, since we derive $\theta\sim272\dg$, thus 
making their claim questionable. The predicted angular separation has just 
passed a minimum and will achieve its next maximum in September 2016 
(\fl{predicted}). Modern instruments should be able to make the 
necessary observations, particularly if \nuo~B is a higher-contrast 
main-sequence star. Our derived mass for the secondary star, if it is 
unevolved, corresponds to a late-K or early-M dwarf having $\mbv\sim8-9$~mag, 
thus \nuo~B may be only six magnitudes or so fainter than \nuo~A. A large 
series of astrometric observations would help better define the entire 
AB orbit, including for instance $\Omega\dbin$ 
which we fixed for our dynamical analyses but may be more inaccurate than we 
suppose. Finally, such observations may also reveal any previously unknown 
stellar companions, such as the hierarchical triple scenario proposed by 
Morais \& Correia (2012).

\bfi
\rotatebox{-90}{\scalebox{0.3}{\includegraphics{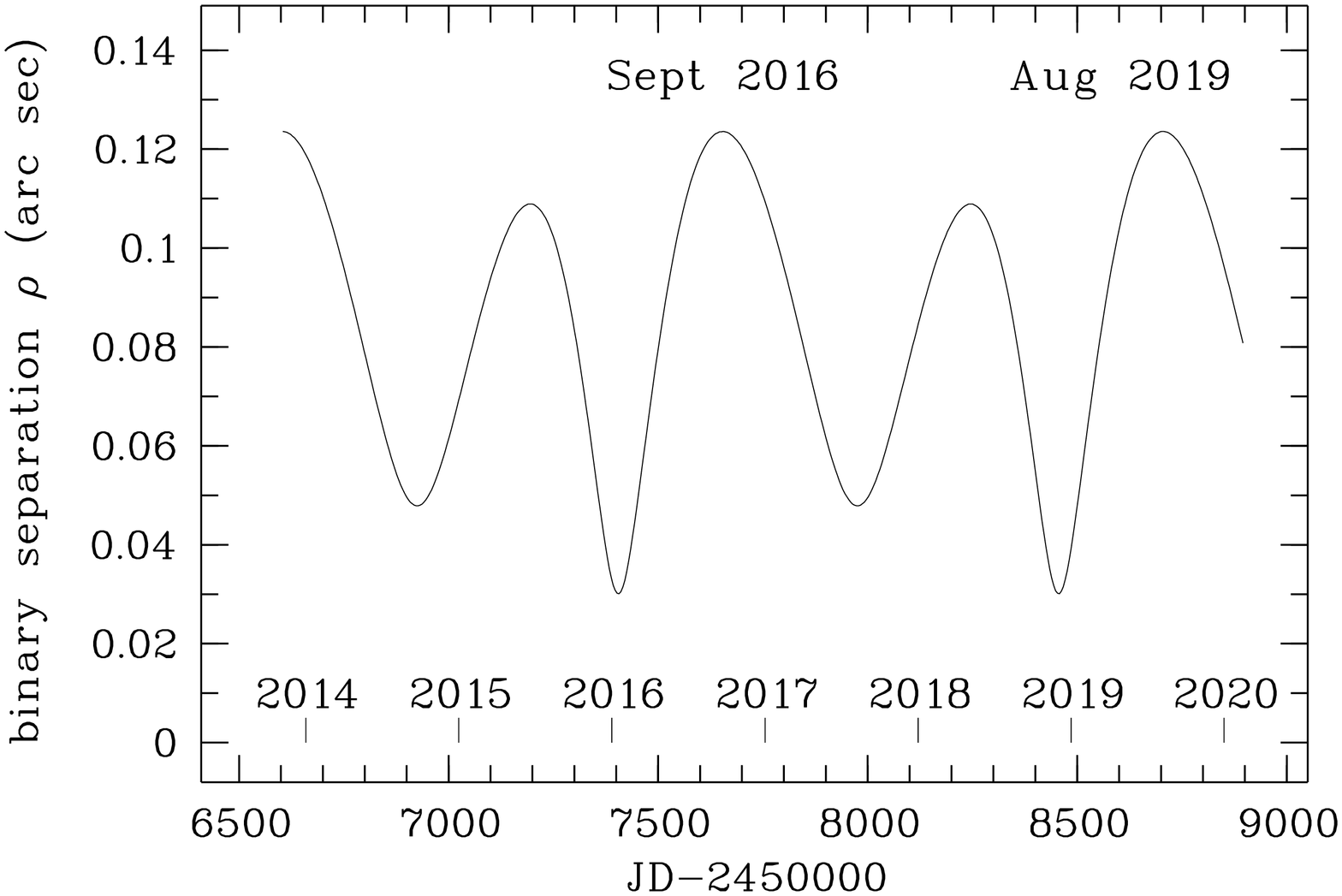}}}
\caption{Predicted separations of the stellar components of \nuo based on 
its parallax (Ramm \oo 2009) and our dynamical orbital elements given in 
\tl{bestmod}.}
\label{predicted}
\efi

\su{Serendipity}
The \hc spectrograph is clearly capable of providing high-precison RVs and has 
been in service for 15 years, yet the \nuo planet is the only one so strongly 
proposed from its spectra. What are the chances such 
an extraordinary planet could be discovered in these circumstances? One would 
imagine they would be very slim indeed. So we ask, just as Wright \oo (2013) 
queried their discovery of the near face-on binary in MARVELS-1, how can we 
make the proposed discovery somewhat more plausible? In the first instance, 
given the Mt.~John telescope's relatively small aperature (1-metre), 
any planet-hunting program is seriously disadvantaged as compared to 
those using much larger apertures in more favourable observing sites. 
Consequently, such programs have rarely been attempted at Mt.~John (but 
see Murdoch, Hearnshaw \& Clark 1993). Instead many \hc-based projects study, 
for instance, pulsating stars and binaries with relatively short periods 
(often less than that of \nuo). These are not ideal or typical targets for 
planet hunting. However \nuo is a suitable object for assessing \hc's 
long-term RV precision, being a very bright, sharp-spectral-lined SB1 (and 
previously having a low-precision orbital solution), is easily circumpolar at 
Mt.~John, and was known to be a very constant star. Hence it was included in 
the first \hc project (Ramm 2004). The $\sim 415$~day RV signal was then 
easily found once it was adequately observed. The serendipitous discovery was 
aided by its far southern declination making \nuo less easily observed at many 
other facilities (though it was initially included in the Pan-Pacific Planet 
Search; Wittenmyer \oo 2011). The remaining surprise may then be limited to 
the fact that something which is presently exceptional and so presumably quite 
rare, is so relatively nearby.

\se{Discussion}
This work demonstrates that the extraordinary retrograde planet in $\nu$ 
Octantis, whilst not yet definitively proven, 
remains the most credible of the known possible scenarios for the RV signal 
first described by Ramm (2004). Credibility for the planet now arises 
primarily from several independent high-precision observational methodologies.
 
The first evidence is the RV signal itself, now demonstrated to be persistent 
for nearly 13 years (2001--2013), and using different strategies for deriving 
the RVs (CCFs and \iod lines). The RMS of our solution for all 1437 RVs is 
8~\ms with $\sqrchi=0.99$ (see \scl{dynsols}). Similar but better long-term RV 
precision has been achieved with the single stars $\delta$ Pav and $\tau$ Cet 
during the parallel observing campaign mentioned in \scl{blue}. For 
instance, the RVs from observations  of $\tau$ Cet ($N=1377$) acquired over 
nearly six years (2008-2014), have an RMS of 4.6~\ms, and the median 
single-night RMS ($N\ge10$ per night) is 2.7~\ms (Bergmann 2015). Hence the 
signal 
cannot be due to instrument errors or the like -- for instance it is also not 
present in RVs derived for the similar SB1 $\beta$ Reticuli (K2IIIb) studied 
in nearly identical circumstances from 2001--2007 (Ramm 2004; Ramm \oo 2009). 

As well as our large new dataset of high-precision \iod RVs, the favourable 
revision of the previously published CCF RVs is expected to have been helpful 
for our dynamical orbit and stability modelling and for future studies. The 
orbits derived are consistent with those reported by Ramm \oo, at least 
with regards the observer-independent elements $a$, $e$ and $P$. These three 
elements are at the heart 
of the controversial claim for the \nuo planet since they place its orbit, on 
average, midway between the stellar components having $\abin\sim2.6$~au.

\su{Luminosity class of $\nu$ Oct A}
The primary star of \nuo is somewhat evolved from the main sequence but we 
wonder if it truly is a giant as is typically claimed  \ie K0III (\eg ESA 
1997) or K1III (\eg Gray \oo 2006), and already questioned by Ramm (2015). 
Cool giants are known to have cyclic RV 
variability periods of the order of hundreds of days and RV 
amplitudes of $50-600\ms$ (Walker \oo 1989; Hatzes \& Cochran 1998; Cummings 
1998). Cool subgiants, whilst less well understood in terms 
of surface dynamics (Hatzes \oo 2003), are generally less dynamically active 
than true giants, so the luminosity class of \nuo~A has some 
significance. The absolute magnitude of \nuo~A 
($\mbv=+2.02$) is maybe as much as two magnitudes fainter than MK-type normal 
giants of luminosity class III, the latter having their locus bounded more or 
less by $-1.5\lesssim\mbv\lesssim+0.5$. The absolute magnitude and spectral 
type suggest \nuo~A is instead no more evolved than luminosity class IIIb - IV 
(see fig.~1 Keenan \& Barnbaum 1999). Also, as well as having many other 
characteristics similar to \nuo~A, the planet-hosting $\gamma$ Cep~A has \mbv 
only 0.5~mag less, and is now classified 
K1\,IV (Fuhrmann 2004). If a re-classification of \nuo A is valid, just as was 
later realised for $\gamma$ Cep~A, our present understanding 
of such less evolved stars is another reason why significant surface dynamics 
would seem less likely (Hatzes \oo 2003; Walker 2012).

\su{Surface dynamics}
Regardless of \nuo~A's true luminosity class, no evidence consistent with any 
kind of significant surface dynamics has been found over several decades and 
at multiple wavelengths \eg in radio (Beasley, 
Stewart \& Carter 1992), microwave (Slee \oo 1989), visible 
(\hip 1992), and X-rays (H\"{u}nsch \oo 1996), and for \caii emission (Warner 
1969; Ramm \oo 2009). P\'{e}rez Mart\'{i}nez, Schr\"{o}der \& Cuntz (2011) 
identify \nuo~A as having a Mg $h+k$ chromospheric emission-line surface flux 
very close to their empirically derived basal flux of 
their sample of 177 cool giants. Our investigations of \caii H-line emission, 
CCF bisectors and line-depth ratios are also all consistent with the 
primary star having negligible dynamical surface behaviour. The LDR and 
bisector results challenge any possibility that surface-activity cycles 
or pulsations are the cause: both here and in Ramm (2015) the extremely 
temperature-sensitive LDRs converted to magnitude differences $\Delta m$ 
strikingly mimic the \hip photometric variations. Furthermore, the predicted 
spot-filling factor $f_{\rm sp}\sim1$\% from the $\Delta m$ values (see Ramm 
2015, fig.~10) can be used to estimate the corresponding RV amplitude (see 
Hatzes 2002). Whilst the predictions of Hatzes were based on solar-type stars, 
since they were apparently applicable for the study of other non-solar-type 
stars \eg HD\,13189 (K2~II; Hatzes \oo 2005), we claim they are equally valid 
here. The predicted amplitude 
($\sim15\ms$) is about $3\times$ smaller than that observed.

Two other details relate to the unlikely possibility of surface dynamics 
being involved. Firstly, as \drb first noted, if the RV signal is related to 
starspots, this would have to be both persistent -- now extending to 12.5 
years -- and geometrically fortuitous to be able to produce the 
near-sinusoidal RV signal. Secondly, the 
estimated period of rotation of \nuo~A $\prot\sim140\pm35$ days, is about one 
third of the RV signal's period. This estimate assumes 
$i_{\rm \sss rot}=\ibin$ which is presumably less likely to be true if a 
significant gravitational event has been involved in this system's history. 
Establishing a reliable and accurate rotational period \prot for the star will 
help decide if starspots have any remaining credibility, though the obvious 
challenge will be achieving this for what seems to be an unspotted and only 
slightly evolved star. The same arguments seem to challenge the 
somewhat more mysterious macroturbulent regions proposed by Hatzes \& 
Cochran (2000) for the RV behaviour of the faster rotating Cepheid-type 
variable Polaris and further discussed in relation to the 
planet-hosting massive K-giant HD~13189 (Hatzes \oo 2005). Pulsations are also 
challenged by the lack of any observational evidence (\eg no cyclic 
photometric or LDR-temperature variations). Finally, 
our Newtonian solutions repeat the finding in Ramm \oo that the orbital 
period of the binary and conjectured planet may be close to a simple 5:2 
ratio. This would be a somewhat unexpected coincidence if the eventual cause 
has a stellar origin.

\su{Hierarchical triple-star scenario}
Our 
orbital solutions also provide new evidence against the possibility that \nuo 
is instead a hierarchical stellar triple. As mentioned in the Introduction, 
Morais \& Correia (2012) presented this as an explanation, claiming the 
tell-tale evidence would be an apsidal precession rate 
$\rm \Delta\omp=-0\dg.86~yr^{-1}$. This is not supported by any orbital 
solutions. We have updated \omp from our full dynamical analysis 
($\omega_{_1}=75.0\pm0.2\dg$; \tl{dynall}). We compared this to $\omega_{_1}$ 
derived from a subset of the 21 historical RVs (Campbell \& Moore 1928; 
Jones 1928). We selected the first thirteen RVs which are confined to 
1904--1911 (JD241\,6640--JD241\,9317) which provide a slightly more 
precise (and hopefully more accurate) \omp than with all 21 RVs. We derived 
$\omp=67.0\pm12\dg$ for the zero-mean-longitude epoch JD241\,7272. A 
single-Keplerian orbit was derived since that is all these few low-precision 
RVs 
can justify. Between this epoch and that defined in \tl{dynall}, the time of 
the first modern RV ($\sim \rm JD245\,2068$), is a time-span of 95 years. The 
apsidal precession Morais \& Correia predict over this 
time-span is about $-82\dg$, which clearly is not the calculated difference 
between these two \omp values. The historical RVs have very low 
precision, but it seems they are easily capable of disclosing such a 
difference if it existed.

To further test the prediction by Morais \& Correia, we fitted a
precessing Keplerian model to the full set of radial velocities. In this
model, we fitted the secondary's orbit as a Keplerian orbit that is allowed
to precess with time, with a precession rate of $\dot{\omega}$. This term
is not governed by any particular physical mechanism such as general
relativity, a stellar quadrupole, or tidal effects. It is merely a free
parameter that we will constrain from the radial velocity data.

Given the large signal-to-noise of the secondary star, we adopted uniform
prior probability distributions in an arbitrarily large range for all of our 
model parameters (P,
K, e, $\omega$, M, $\dot{\omega}$, etc.). We performed an affine-invariant
MCMC (Foreman-Mackey et al. 2013) to sample from the posterior
distribution using a likelihood function similar to as described in
Nelson et al. (2014a). After burning-in sufficiently, we sampled from the
Markov chains and constrained $\dot{\omega}=+0.00951\pm0.00006$ deg per
year. Not only is this result two orders of magnitude off from the
prediction, but it is also in the opposite direction (\ie a prograde
precession).

\su{Stability modelling}
Our stability studies have provided mixed results for credibility for the 
planet. On one hand, the observations provided very precise orbital and 
mass constraints for a retrograde planet model. However, none of the posterior 
samples were dynamically stable for $10^6$ years. 
One might suggest we are actually witnessing a system on the brink of orbital 
destabilization, but given the time-scale of instability compared with the age 
of the star, this would be highly improbable.
Several trial runs of the observations revealed many local minima which 
could easily trap our Markov chains depending on our set of initial conditions.
It is possible, therefore, that our results are based on samples from a local 
rather than the true global minimum, which, based on the results of 
Go{\'z}dziewski et~al. (2013), would not be too surprising.

The grid search revealed a region of parameter space with a higher probability 
of long-term stability. Only a relatively small fraction 
of models survived to $10^7$ years ($\sim 5$\%). This may, however, be 
anticipated given the extreme nature of the system we are modelling. The 
majority of the stable models are somewhat distant from the best-fitting 
solution (see \fl{stabilities}), but one model whose grid values are all 
within 1-$\sigma$ of the best fit (see \tl{bestmod}) was stable to at least 
35~Myrs years 
for our grid's four time-steps, and stable beyond $10^8$ years for two of them 
($\tau=0.4$ and 11 days). If the conjectured system's formation arose 
from a short-lived gravitational event (\eg a passing star) the mutually 
inclined orbit, rather than a coplanar one, would seem to be more consistent 
with that possibility. A hypothesis of a moderate mutual inclination could 
also provide other significant advantages for more quickly resolving the true 
nature of the enigmatic \nuo system \eg in the form of 
verifiable predictions for orbit evolution of the binary and/or the 
conjectured planet such as Kozai-like mechanisms (Kozai 1962).

\se{Conclusion}
To date, the observational evidence more strongly supports the retrograde 
planet's reality than any other known alternative. Yet, due to its exceptional 
nature, definitive independent evidence is essential for deciding the fate of 
this controversial planet. Spectra acquired 
with much higher $S/N$ and resolving power (and overall fidelity) will help 
decide the credibility of any stellar-origin scenarios, and additional 
preferably more precise RVs will better define the 
conjectured orbit and facilitate better predictions of its evolution.\fn{See 
\eg Fischer \oo (2016) for the state-of-the-art of RV precision and its future 
goals and challenges. 
The work of Wright \& Eastman (2014) identifies one avenue of review of our 
RVs based on the barycentric corrections, which we have no reason to suspect 
are significantly inaccurate, but which we have provided for this purpose.} 
This can lead to a re-investigation of a self-consistent Newtonian model that 
also incorporates the \hip and future astrometry. If the 
planet's existence is confirmed, a theoretical explanation will be 
needed. A gravitational interaction with a single or multiple bodies would 
presumably require the least, if any, revision to our present understanding of 
such scenarios. Some other migration process involving the stars or their 
planetary companion/s may also be applicable, though 
as with all formation scenarios, this too would seem to be a challenging one. 
If these options proved not to be viable, then \nuo would instead aggressively 
confront our present 
understanding of coeval planet formation in such tight binaries, regardless of 
the system's original configuration. For many, the latter would be a very 
difficult scenario to believe. It is perhaps, however, prudent to recall that 
not so long ago (the 1980s and even early 1990s) it would have been almost as 
hard to believe hot-Jupiter exoplanets existed either, or indeed that there 
was any sense looking for exoplanets anywhere (Walker 2012). 

Evidence against the planet should be as equally robust as evidence for it, 
since, if it is real but wrongly dismissed, we would presumably loose or at 
least delay our many opportunities for the study of its unique 
characteristics. If the planet is non-existent, but all the astrometric, 
spectroscopic and photometric evidence is accurate, the consequences quite 
likely exceed 
that arising from the planet's reality. We would have to acknowledge there 
is a new variation of stellar behaviour or orbit-related process that is 
consistent with all the current evidence. Such a discovery would have 
far-reaching consequences, both as a 
discovery in itself, but as a significant challenge to the existence of many 
other already {\it and} yet-to-be claimed RV-detected planets, perhaps 
regardless of the host star's evolution.

\se{Acknowledgements}
The Mt John programme was generously funded by Marsden Grant UOC 1007 
administered by the Royal Society of New Zealand. The project was also 
supported in part by the Australian Research Council Discovery Grant 
DP110101007 (RW \& FG). The authors thank McDonald Observatory, University of 
Texas at Austin for sharing the Sandiford iodine cell. The authors acknowledge 
the contribution of the 
New Zealand computer and analytics services NESI (New Zealand eScience 
Infrastructure), funded jointly by NeSI's collaborator institutions and 
through the Ministry of Business, Innovation and Employment 
(URL http://www.nesi.org.nz). DJR thanks Fran\c{c}ois Bissey for his support 
using these High-Performance Computing facilities at UC. We thank Stuart 
Barnes for his enthusiasm establishing the early work of this 
iodine-cell campaign and Jovan Skuljan for reconfiguring his 
reduction software HRSP for the larger \kfour detector. BEN thanks Eric Ford, 
Jason Wright, and Trifon Trifonov for useful 
conversations about various interpretations of the data. DJR also thanks 
Gerald Handler for helpful discussions relating to pulsating 
stars, Trifon Trifonov for identifying a data anomaly and other useful 
comments, and Aleksandra Jarmolik for many helpful discussions including, on 
the lighter side, her recommendation to read Stanis{\l}aw Lem's ``Solaris'', 
it seems related to this work on several levels. We also 
appreciated the support of all staff at the Mt.~John Observatory. Our 
work has benefited from the databases provided by SIMBAD, VizieR (CDS, France) 
and the NASA ADS.

\lab{lastpage}
\end{document}